\documentclass[prb, showpacs,amstext,amsxtra,amssymb,latexsym,bbm,twocolumn]{revtex4}
\usepackage{graphicx,epsf}
\usepackage{amsmath,float}
\usepackage{amsfonts}
\usepackage{amsthm}
\usepackage{amssymb}
\usepackage{multirow}
\usepackage{color}
\usepackage{verbatim}
\newcommand{\BB}{{B}}
\newcommand{\BBS}{{B_f}}

\renewcommand{\k}{{\boldsymbol k}}

\newcommand{\be}{\begin{equation}}
\newcommand{\ee}{\end{equation}}

\newcommand{\Q}{{\boldsymbol Q}}

\newcommand{\ep}{\epsilon}

\def\t6as{$\mathrm{(TMTSF)_{2}AsF_{6}}$}

\def\tmc{$\mathrm{(TMTSF)_{2}ClO_{4}}$\,}
\def\tms{$\mathrm{(TMTSF)_{2}AsF_{6(1-x)}SbF_{6x}}$}\,

\def\tmttfsbf6{$\mathrm{(TMTTF)_{2}SbF_{6}}$\,}
\def\tmttfasf6{$\mathrm{(TMTTF)_{2}AsF_{6}}$\,}
\def\tmttfbf4{$\mathrm{(TMTTF)_{2}BF_{4}}$\,}
\def\tmtsfreo4{$\mathrm{(TMTSF)_{2}ReO_{4}}$\,}
\def\tmno3{$\mathrm{(TMTSF)_{2}NO_{3}}$\,}
\def\tm2x{$\mathrm{(TM)_{2}X}$\,}

\,
\,

\def\hc2{$H_{c2}$\,}

\def\tmp6{$\mathrm{(TMTSF)_{2}PF_{6}}$\,}
\def\tms2x{$\mathrm{(TMTSF)_{2}X}$}
\def\tm2x{$\mathrm{(TM)_{2}X}$\,}
\def\as{$\mathrm{AsF_{6}}$}

\def\pf{$\mathrm{PF_{6}}$}
\def\re{$\mathrm{ReO_{4}}$}

\def\cl{$\mathrm{ClO_{4}}$}
\def\no{$\mathrm{NO_{3}}$}
\def\4fb{$\mathrm{BF_{4}}$}

\def\tsx{$\mathrm{(TMTSF)_{2}X}$\,}

\def\reo4{$\mathrm{ReO_{4}}$}
\def\bedtttfreo4{$\mathrm{(BEDT-TTF)_{2}ReO_{4}}$\,}
\def\et2i3{$\mathrm{(ET)_{2}I_{3}}$\,}
\def\et2x{$\mathrm{(ET)_{2}X}$\,}
\def\ket2x{$\mathrm{\kappa-(ET)_{2}X}$\,}

\def\ket2x{$\mathrm{\kappa-(ET)_{2}X}$\,}

\def\betsfecl4{$\mathrm{(BETS)_{2}FeCl_{4}}$\,}

\def\et{$\mathrm{ET}$\,}

\newcommand{\red}{\color{red}}

\begin{document}


\title{Rapid magnetic oscillations  and magnetic breakdown in quasi-1D conductors}
\author{G. Montambaux and D. J\'erome}
\affiliation{Laboratoire de Physique des Solides, CNRS UMR 8502, Univ. Paris-Sud, F-91405 Orsay Cedex, France}
\date{October 12, 2015}


\begin{abstract}
We review the physics of magnetic  quantum oscillations in quasi-one dimensional conductors with an open Fermi surface, in the presence of modulated order. We emphasize the difference between situations where a modulation couples states on the {\it same} side of the Fermi surface and a modulation couples states on {\it opposite} sides of the Fermi surface. We also consider   cases where several modulations coexist,   which may lead to a complex reorganization of the Fermi surface. The interplay between nesting effects and magnetic breakdown is discussed. The experimental situation is reviewed.
\end{abstract}
\maketitle
\section{Introduction}

It is well known that magnetic oscillations   in thermodynamic and transport properties  originate from the Landau quantization of {\it closed} electronic orbits. The existence of such oscillations in quasi-1D conductors with an {\it open} Fermi Surface (FS), especially studied in compounds of the Bechgaard salts family, has thus been a long standing problem (for a review, see Refs.[\onlinecite{Lebedbook,Yamajibook}]).
Various mechanisms have been invoked to explain the existence of these quantum oscillations in the presence of open orbits. Most of them are based on the existence of an  external periodic potential which permits a modification of the Fermi surface.  Other mechanisms like the magnetic field modulation of electron-electron scattering\cite{eescattering}  and the rich physics of the angular oscillations are not discussed here.\cite{Yamajibook}

\medskip

One of these mechanisms is the Density Wave (DW) ordering due to almost perfect nesting of the Fermi surface (FS). Such ordering leaves small {\it closed}
  electronic pockets of unpaired carriers the  size of which is related to the deviation from perfect nesting, as recalled   later in this paper (Fig. \ref{fig_QN}). In a magnetic field  $\BB$ applied along a direction perpendicular to the most conducting planes, the quantization of the electronic motion along these closed pockets leads to Shubnikov-de Haas (SdH) oscillations  (periodic in $1/\BB$) the period of which is proportional to the size of the pocket.
 The typical field $\BBS$ characteristic of the oscillations is proportional to the area ${\cal A}$ of the closed orbits in reciprocal space $\BBS ={\hbar \over 2  \pi e} {\cal A}$. The characteristic energy of deviation from perfect nesting, named $t'_b$, is usually of order of $10$-$30$\,K, so that ${\cal A}   \propto {  t'_b \over \hbar   v_F b}$ and the typical field $\BBS \propto {t'_b \over e   v_F b}$ is of order of a few dozen teslas. In Bechgaard salts, the competition between Spin Density Wave (SDW) ordering and the quantization due the  magnetic field   leads to a cascade of SDW subphases in which the  Hall effect is quantized.\cite{Lebedbook,Yamajibook,GL,FISDW,Yakovenko}
\medskip

 In this paper,  we   focus  our study to the understanding  of  the so-called
 {\it Rapid Oscillations} (RO), described by a much larger characteristic field, of order of few hundred teslas, believed to be related to the typical warping of the FS  related to an energy scale $t_b \gg t_b'$, therefore to much larger orbits the  existence of which cannot be explained by DW ordering alone   (${\cal A}   \propto {  t_b \over \hbar   v_F b}$ and $\BBS \propto {t_b \over e   v_F b}$).
\medskip

 We examine various kind of external periodic potentials which may give rise to such rapid oscillations.
 We consider a  simple band model in order to  study the effect of different modulations on the electronic spectrum and their consequence on the structure of the   magnetic field induced quantum oscillations.
We start from the widely used two-dimensional tight-binding model describing a metallic phase with a  simple orthorhombic dispersion with hopping parameters  $t_a$ along the $x$ direction  of the conducting chains and $t_b$ along the perpendicular $y$ direction. The magnetic field is applied along the $z$ direction   (the $c^*$ direction in   Bechgaard salts having triclinic symmetry). The dispersion relation  may be linearized along the high conductivity direction and the  modulation along the transverse direction is then described by two harmonics with amplitudes $t_b$ and $t'_b$~:\cite{Lebedbook,Yamajibook,GL,FISDW}

\be \ep_\k= \ep_F+ \hbar v_F (|k_x| - k_F) - 2 t_b \cos k_y b -2 t_b'\cos 2 k_y b \ .   \label{DR1}  \ee
We take the Fermi energy $\ep_F= 2 t_a \cos k_F a$ as the origin of the energies and the Fermi velocity $v_F$ is given by $\hbar v_F= 2 t_a  a \sin k_F a$. The corresponding FS is made of two warped sheets located at $\pm k_F$ (Fig. \ref{fig_Qperp}-a).
The amplitude of the warping is given by $t_b$.  As we will recall in section \ref{sect:nesting}, a wave vector $\Q_N=(2 k_F,\pi/b)$ almost perfectly nests the two sheets.\cite{remarknesting} The deviation from perfect nesting is then related to the amplitude $t'_b$.
\medskip

In this paper, we emphasize the  possible  existence of two different kinds of  periodic structural modulations and their consequences on the structure of the FS and on  the nature of the magnetic oscillations~:

i) Modulations with wave vector  along the transverse direction to the  conducting chains (Fig. \ref{fig_Qperp}). A modulation at wave vector $\Q_\perp=({0, \pi/b})$ couples electronic states located on the {\it same} side of the FS. In a magnetic field, two {\it open } trajectories flow along the   {\it same direction}  and may interfere at special positions in reciprocal space, realizing a double path interferometer.\cite{Stark}

ii)   Modulations the  wave vector of which has a $2 k_F$ component which couples states located on {\it opposite} sides of the FS (Fig. \ref{fig_Qparallel}). A modulation at wave vector $\Q_\parallel=({2 k_F, 0})$ opens a gap at a transverse position $\pm \pi/(2b)$ and leaves  {\it closed} orbits  of size proportional to the energy scale $t_b$. Quantization of these closed orbits in a magnetic field leads to Shubnikov-de Haas (SdH) oscillations the  frequency  of which is proportional to $t_b$.

As shown on Figs.\ref{fig_Qperp},\ref{fig_Qparallel}, the dynamics in a magnetic field is quite different between these two cases, since in the first case, the coupled trajectories have the {\it same} direction in a magnetic field while in the second case they follow {\it opposite} directions.
\medskip

  The present work is motivated by several puzzling experiments showing rapid oscillations (frequency $t_b$) performed in Bechgaard salts. The next section presents a brief overview of these experiments. Then we consider different different situations corresponding to different modulations. The main goal of the present paper is not to address in detail a given experiment but rather to show the variety of the possible mechanisms. Whenever it looks appropriate, we   refer to a given experimental result.   The outline  is the following. In section \ref{sect:transverse},
 we consider a transverse modulation   ($\Q_\perp$)  which connects states on the same side of the FS. This modulation induces a pair of trajectories which may interfere in a magnetic field through Magnetic Breakdown (MB).\cite{Stark} This interference effect is reminiscent of the St\"uckelberg oscillations between  two Landau-Zener transitions.\cite{Shevchenko} In section  \ref{sect:longitudinal}, we consider a longitudinal modulation  ($\Q_\parallel$)  which naturally produces closed orbits of the appropriate size to induce rapid quantum oscillations. For these two cases, we calculate explicitly the variation of the characteristic field $\BBS$ with the amplitude of the gap induced by the modulation in the electronic spectrum. Section \ref{sect:nesting} recalls the case of almost perfect nesting induced by a DW  ($\Q_N$), which leads to small pockets and slow oscillations. Then we consider    situations where {\it two modulations coexist}  ($\Q_N$ and $\Q_\perp$ in section  \ref{sect:nesting-transverse}; $\Q_N$ and $\Q_\parallel$ in section  \ref{sect:nesting-longitudinal}). Such a coexistence leads to a more complex structure of the Fermi pockets in the ordered phase.  A similar mechanism may occur in a triclinic crystal where the two sheets are translated with respect to each other, so that naturally two  DWs may coexist. This is discussed in section \ref{sect:nesting-commensurate}. We then conclude on the experimental situation.

 \section{Experimental overview}
\label{sect:experiments}

  We start with an overview of the experiments showing quantum oscillations in   the quasi-1D conductors and restrict ourselves to the members of the  Bechgaard salts \tsx \, family and describe the  two  kinds of oscillatory behaviors (periodic in $1/\BB$) that  have been observed.

 In \tmc, \cite{Ribault83,Chaikin83,Ulmet86} \tmp6,\cite{Kwak81,Ulmet85}  and \tmtsfreo4  under pressure,\cite{Kang91b} oscillations  with a frequency around $30$  T  (the so-called {\it slow oscillations}) are  observed at low temperature when the metallic phase  is stable {\it{albeit}}
above a threshold magnetic field  $\approx$ 5-8 T.\cite{Takahashi82,Ulmet83}
  These oscillations are now fairly well understood in terms of  the stabilization of  field-induced spin density wave phases (FISDW)\cite{FISDW} and will not be discussed in this paper.

Quite an  intriguing feature is the observation of oscillations  with a much higher characteristic frequency, typically $250$-$300$ T, the so-called {\it Rapid  Oscillations} (RO)  in the ambient pressure SDW phase of \tmp6 \,  \cite{Ulmet85} and \t6as,\cite{Ulmet97a,Brooks99}  in the high magnetic field   ($N=0$) FISDW of \tmc \cite{Uji96} and \tmp6 under pressure,\cite{Kornilov07}  and even in the   SDW phase   of rapidly cooled (quenched~: Q)    Q-\tmc at ambient pressure.\cite{Brooks99} They are also observed  in the metallic phase of  slowly cooled (relaxed~: R)  R-\tmc at ambient pressure,\cite{Ulmet86,Uji96,Uji97}
and \tmtsfreo4 under pressure.\cite{Kang91b}

 The  salt   \tmno3 is somewhat peculiar since slow  and rapid oscillations are observed in the ambient pressure SDW phase  under low and high fields respectively.\cite{Audouard94}
When the SDW phase is suppressed under a pressure exceeding $8$ kbar,\cite{Mazaud80} RO are the only oscillations to survive.\cite{Kang95,Vignolles05}

  From these observations, we may draw three important conclusions. i)
   The analysis of RO in all four compounds \cl, \pf, \re, \no\, show that their existence is not   necessarily related  to the FISDW phases. ii)
The  frequency of these rapid oscillations is related to  interchain coupling $t_b$, that is to the warping of the open Fermi surface. iii) In several cases the temperature dependence of the amplitude exhibits a marked deviation  from the conventional Lifshitz-Kosevich description, especially a sudden vanishing of the oscillations at low temperature.\cite{Brooks99} Guided by these observations, we now propose an overview of all situations where RO arise in these materials  with an unified theoretical model based on the experimental results.

\section{Transverse $(0,\pi/b)$ modulation}
\label{sect:transverse}

\begin{figure}[h!]
\begin{center}
{\epsfxsize 4cm \epsffile{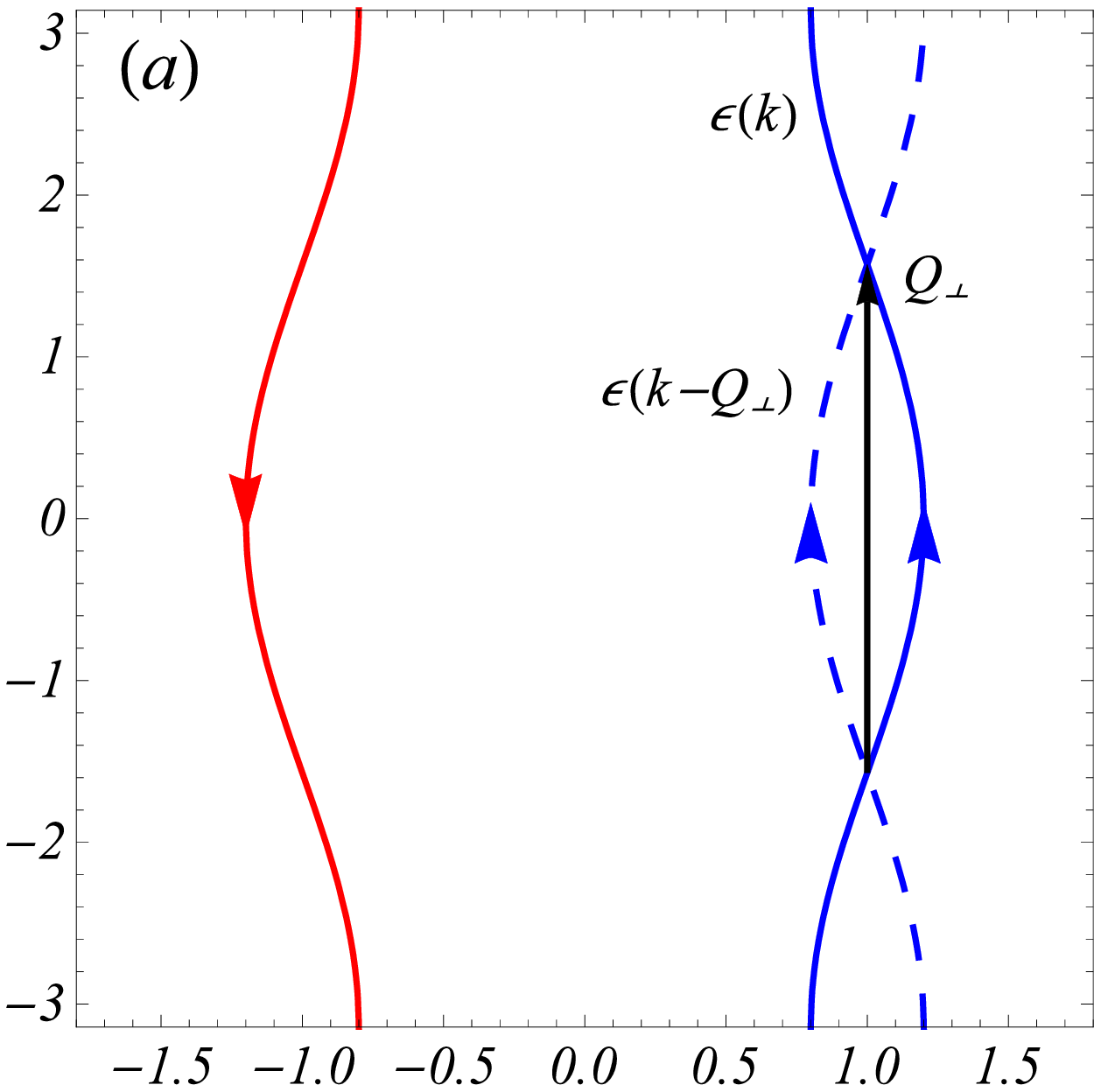}}
{\epsfxsize 4cm \epsffile{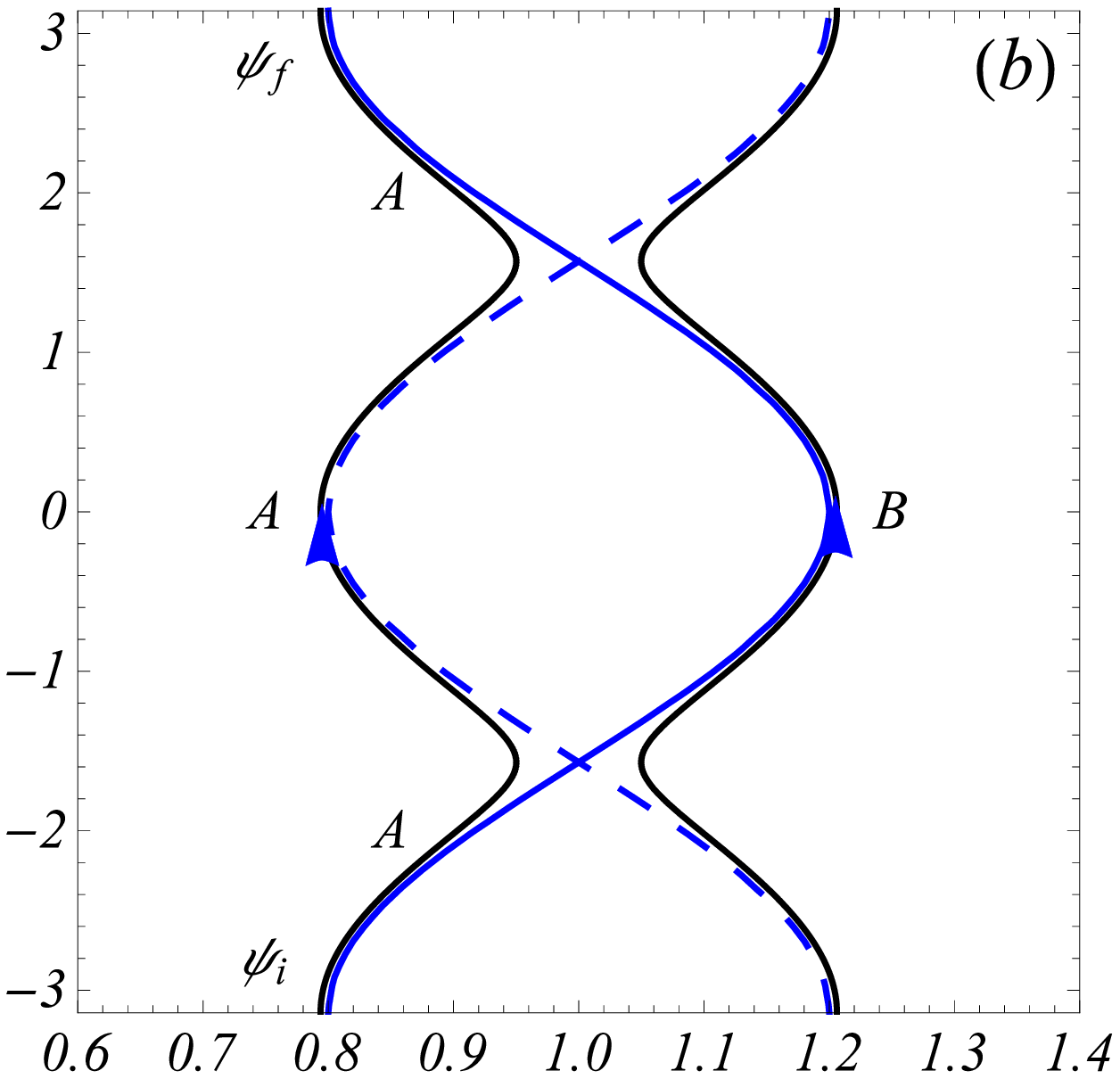}}
\caption{\small \it a) A modulation at a transverse wave vector $\Q_\perp=(0,\pi/b)$ couples states on the same side  of the FS. In a magnetic field, the coupled trajectories (blue and dashed blue) flow along the same direction. b) The opening of a gap creates two open warped sheets of the FS (black), with  the possibility magnetic breakdown at $(k_F, \pm \pi/(2b))$. An electron initially on the sheet A  may travel along two interfering paths (AAA and ABA), leading to magnetic oscillations in the conductance, with a frequency proportional to $t_b$.}
\label{fig_Qperp}
\end{center}
\end{figure}

We first consider the existence of a transverse modulation with amplitude $\Delta_\perp$ at wave vector $\Q_\perp=(0,\pi/b)$, as could be induced by an anion modulation along the transverse $y$ direction created by the ordering of ClO$_4$ anions  in (TMTSF)$_2$ClO$_4$. The modulation couples states on the {\it same} side of the FS (Fig. \ref{fig_Qperp}).

In the presence of a magnetic field $\BB$, the electrons experience a motion along an open FS and quantum oscillations are usually not expected for an open FS. However, the situation is different here since {\it two} open trajectories run at short distance in $\k$ space and magnetic breakdown near the points $(k_F, \pm \pi/2b)$  is possible.\cite{Stark,Uji96,Uji2}

The potential  $\Delta_\perp$  couples $| \k \rangle $ and $|\k - \Q_\perp \rangle$, that is states on the {\it same} side of the Fermi surface. For $k_x >0$, this coupling is described by the effective Hamiltonian
\be{\cal H}_A(\k)= \left(
  \begin{array}{cc}
    \ep_{\k} & \Delta_\perp \\
    \Delta_\perp & \ep_{\k-\Q_\perp} \\
  \end{array}
\right) \ee
with $\ep_{\k}$ given by (\ref{DR1}) and
\be \ep_{\k - \Q_\perp} =  \hbar v_F ( k_x - k_F) + 2 t_b \cos k_y b - 2 t_b' \cos 2 k_y b  \ . \label{shiftedQA}  \ee
The $t'_b$ term only slightly distorts the FS, but does not change the physics at all.  Therefore we set here $t'_b=0$.
The new spectrum is given by
\be E_{\k}= \hbar v_F ( k_x  - k_F)    \pm \sqrt{ \Delta_\perp^2 + 4 t_b^2 \cos^2 k_y b } \label{EkA} \ee
and   the equation of the corresponding FS ($E_\k=0$) is
\be   k_x  =  k_F     \pm {1 \over \hbar v_F} \sqrt{ 4 t_b^2 \cos^2 k_y b +  \Delta_\perp^2  } \ .  \label{FSA} \ee
It is shown on Fig.\ref{fig_Qperp}. It consists in two warped sheets along the same side of the FS.
\medskip

\subsection{Open orbits and magnetic breakdown}

We estimate now the probability of  magnetic breakdown in the vicinity of the gap separating these two sheets. Near the Bragg reflexion $k_y =\pm \pi/(2b)$, and expanding $k_y=\xi \pi/(2b) + q_y$ with $\xi= \pm 1$,  the Hamiltonian  has the form~:

\be{\cal H}_A(q_y)=   \left(
  \begin{array}{cc}
   2 \xi  t_b b  \, q_y  & \Delta_\perp \\
    \Delta_\perp  & -2 \xi t_b b \,  q_y  \\
  \end{array}
\right) \ee
with the spectrum
\be E_{\k}= \pm \sqrt{ \Delta_\perp^2 + 4 t_b^2 b^2  q_y^2  } \ .  \ee
We define the transverse velocity as $\hbar v_y = 2 t_b b$.
In a magnetic field $\BB$, the transverse wave vector $q_y$ varies linearly with the field, due to the Lorentz force $F= e v_F \BB$
\be \hbar q_y = e v_F \BB t \ee
so that the time dependent Hamiltonian simply reads

 \be{\cal H}_A(t)=    \left(
  \begin{array}{cc}
  \xi  v_y F t   & \Delta_\perp \\
    \Delta_\perp & - \xi v_y F t  \\
  \end{array}
\right) \ .  \ee
This problem is exactly equivalent to the Landau-Zener problem associated with the one-dimensional adiabatic spectrum\cite{LZ}
\be E(t)= \pm \sqrt{ \Delta_\perp^2 +v_y^2 F^2 t^2  }  \ . \ee
 It is known that in this case, the Landau-Zener probability is therefore given by  (the gap being $2 \Delta_\perp$)~:\cite{LZ}
\be \displaystyle p_\perp \equiv e^{- 2 \pi \delta} = e^{\displaystyle -\pi  {\Delta_\perp^2 \over \hbar  v_y F} } \ . \ee
$\delta$ is called the adiabaticity parameter. In our case, the MB probability is given by

\begin{equation}
\displaystyle
p_\perp= e^{\displaystyle -{\BB_\perp \over \BB} }  \quad     \mbox{with} \quad  \BB_\perp= \pi  {\Delta_\perp^2   \over  e \hbar  v_y  v_F }= {\pi \over 2}{ \Delta_\perp^2 \over e t_b b v_F } \ .
\label{MB1} \end{equation}
which is of the form found in Schoenberg (eq. 7.13).\cite{Schoenberg,Blount}
At this stage, it is useful to compare this result with similar but different formulas used in the literature for $p_\perp$.
In Ref.[\onlinecite{Uji96}], Uji {\it et al.} address the rapid oscillations in     (TMTST)$_2$ ClO$_4$. The RO in the metallic phase are attributed to the Stark-St\"uckelberg mechanism that we discuss below. The characteristic field for magnetic breakdown is evaluated as
$ \BB_\perp = {\Delta_\perp^2  m_c  \over    \hbar  e \ep_F    }$
where $m_c$ is a cyclotron effective mass defined as $m_c= \hbar /(e v_F b)$ and assumed to be of the order of the free electron mass. Our result disagrees with this estimate since the energy scale in the denominator is proportional to $t_b$ and not to $\ep_F \propto t_a$.
In
  Ref.[\onlinecite{Uji2}],  a slightly more refined  formula  $\BB_\perp = {\Delta_\perp^2  m^*  \over    \hbar  e \ep_F  \sin(2 \theta)   }$ is in qualitative agreement with our result since
 $\theta$ is defined as the scattering angle at the gap, and is therefore of order $t_b/t_a$.

\subsection{Stark-St\"uckelberg oscillations}

 Due to the modulation at wave vector $\Q_\perp$,  one side of the FS is now made of two open sheets. In a magnetic field, the electrons travel   on both sheets along the {\it same direction}, and possibly experience a tunneling through magnetic breakdown from one sheet to the other. Since this tunneling   occurs in two different places (Fig. \ref{fig_Qperp}-b), the contributions corresponding to the two different paths may interfere. This phenomenon occurring between two LZ transitions is known as St\"uckelberg oscillations.\cite{Stuckelberg} In the context of electronic magnetic breakdown, it has been proposed by Stark {\it et al.} to explain quantum oscillations in Mg,\cite{Stark} and by Uji {\it et al.} do interpret the rapid resistance oscillations in the Bechgaard salt (TMTST)$_2$ ClO$_4$.\cite{Uji96,Uji97}    We present here a quantitative picture of this effect.

Consider an electron in a magnetic field along one open sheet of the FS. During one period along the BZ, it experiences two LZ transitions to the neighboring sheet (Fig. \ref{fig_Qperp}-b). The tunnel    probability amplitude $\sqrt{p_\perp}$   has been calculated above (\ref{MB1}). The probability amplitude to stay on the same band is $\sqrt{1- p_\perp}$.  Therefore calling $\psi_i$ the wave function   on one sheet  at one end of the BZ, the wave function $\psi_f$   on the same sheet  at the other end is~:
\be \psi_f= [p_\perp e^{i \phi_B} +(1-p_\perp) e^{i(\phi_A - 2   \varphi_s)} ]\,  \psi_i  \ .  \label{amplitudeLZ} \ee
The first term corresponds to two "transmissions" from one sheet (A, see Fig. \ref{fig_Qperp}-b) to the neighboring one (B), the second term corresponds to two reflections to the initial sheet (A). The phase  $\phi_{A,B}= {1 \over \hbar} \int E_{A,B}(t) dt$ is the dynamical phase along the path $A,B$ between the two MB events. The phase $\varphi_s$ depends on the adiabaticity parameter $\delta$, therefore on the amplitude of the magnetic field. It is   the so-called Stokes phase   accumulated at a Landau-Zener reflection~:
   $\varphi_s= \pi/4 + \delta ( \ln \delta -1) + \mbox{arg} \Gamma( 1 - i \delta)$.\cite{Shevchenko} Here, we have   $2 \pi \delta=  \BB_\perp / \BB$. Therefore $  \varphi_s$ varies between $0$ in the adiabatic limit ($\delta \rightarrow \infty$, absence of magnetic breakdown,   $\BB \ll \BB_\perp$) to $\pi/4$ in the diabatic limit ($\delta \rightarrow 0$, strong magnetic breakdown,   $\BB \gg \BB_\perp$).
From (\ref{amplitudeLZ}), the probability for the electron to stay on the same sheet after one period is therefore given by~:
\begin{eqnarray}
 |\psi_f|^2/|\psi_i|^2&=& p_\perp^2 + (1-p_\perp)^2  \\
 &+& 2 p_\perp (1-p_\perp) \cos( \phi + 2 \varphi_s)  \end{eqnarray}
where $\phi = \phi_B - \phi_A$ is the magnetic field dependent dynamical phase accumulated between the two paths.
It is given by
$\phi= {1 \over \hbar} \int_{-t_0}^{t_0} \Delta E(t)  dt $ with $\Delta E(t)= 2 \sqrt{\Delta_\perp^2  + 4 t_b^2 \cos^2 {e v_F b \BB t \over \hbar}}$ and $e v_F b \BB  t_0 /\hbar = \pi/2$. In the limit, $\Delta_\perp \ll t_b$, the dynamical phase is simply given by $\phi(0) = {8 t_b \over e v_F b \BB}$.   It {\it increases} with $\Delta_\perp$ as
$\phi(\Delta_\perp) = \phi(0) F_\perp(\Delta_\perp/2 t_b)$ where the function $F_\perp(x)$ is given by
\be
F_\perp(x)= \int_0^{\pi/2} \sqrt{\cos^2 t + x^2}\,  dt  \ . \ee

\begin{figure}[h!]
\begin{center}
\epsfxsize 8cm \epsffile{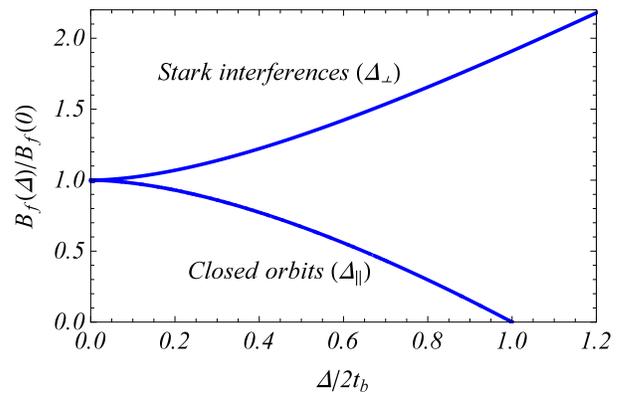}
\caption{\small \it   Gap dependence of the characteristic field $\BBS$ of the rapid oscillations, for a transverse modulation $\Delta_\perp$ leading to Stark interferences
 and for a longitudinal modulation $\Delta_\parallel$ leading to closed quantized orbits (see Eqs.\ref{Bstarperp}, \ref{Bstarparallel}).   This corresponds to the situation respectively encountered in (TMTSF)$_2$ClO$_4$ and (TMTSF)$_2$ReO$_4$ under pressure on the one hand, and (TMTSF)$_2$NO$_3$  at ambient pressure  on the other hand, as discussed in the summary section (\ref{sect:conclusion}).}
\label{fig:freqs}
\end{center}
\end{figure}

This interference mechanism leads to oscillations of the conductivity of the form~:\cite{Stark}

\be \sigma_{osc}  \propto 2 p_\perp (1-p_\perp) \cos ( 2 \pi {\BBS \over \BB} +  2   \varphi_s)  \ ,  \ee
where the
  the characteristic field $\BB^*$ is
\be \BBS(\Delta_\perp) = {4 t_b \over \pi e v_F b}\ F_\perp\left( {\Delta_\perp \over 2 t_b}\right) \ .   \label{Bstarperp} \ee
It increases with $\Delta_\perp$ since the distance between interfering orbits increases in $\k$ space. It is plotted in Fig. \ref{fig:freqs}.

As already explained in Refs.[\onlinecite{Stark,Uji96,Uji2}], these oscillations are only visible in transport and they ressemble
Shubnikov-de Haas   oscillation,   however with a different temperature dependence. Since the dynamical phase is energy independent (it does not depend on the position of the Fermi energy), there is no thermal damping of the  oscillations.  Their temperature dependence  is due to that of the scattering time.\cite{Uji96}

\begin{figure}[h!]
\begin{center}
\epsfxsize 8cm \epsffile{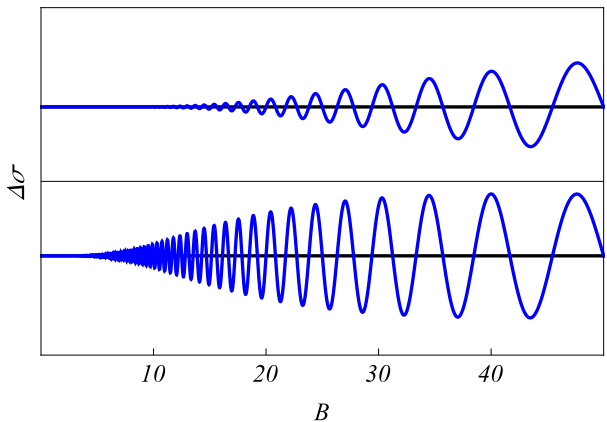}
\caption{\small \it   Evolution of the Stark oscillations in a magnetic field (arbitrary units). Here we have taken $\BBS=250$ T, $\BB_\perp=70$ T (top curve) and $\BB_\perp=30$ T (bottom curve).}
\label{fig:stark-oscillations}
\end{center}
\end{figure}

Since these quantum oscillations involve two LZ transitions, they vanish for a large gap $\Delta_\perp$ ($p_\perp \rightarrow 0$) and also for small gap  ($p_\perp \rightarrow 1$). They vary as~:

\be \sigma_{osc}  \propto   e^{-\BB_\perp/\BB} \left( 1- e^{-\BB_\perp/\BB} \right)\cos \left( 2 \pi {\BBS \over \BB} + 2 \varphi_s \right) \ .  \ee
Typical variations are shown on Fig.\ref{fig:stark-oscillations}.   They are maximal for $p_\perp=1/2$, that is for a field $B_m=B_\perp/\ln 2$.
  This mechanism has been proposed as a possible explanation for the rapid oscillations observed in the metallic phase of (TMTSF)$_2$ClO$_4$.\cite{Uji96,Uji97}
   With the known physical parameters in this salt, $a=3.65 \AA$, $b=7.7 \AA$, $t_a \simeq 3 000$ K, $t_b \simeq 3 00$ K, and a recent estimate\cite{Alemany14} of the anion gap $\Delta_\perp \simeq 14$ meV, one finds an order of magnitude   $B_\perp \simeq 40$ T which is compatible with experiments. The same mechanism should also explain the RO in  (TMTSF)$_2$ReO$_4$ which seem to be of the same nature.\cite{Kang91b}
   For completion, let us mention   ref.[\onlinecite{eescattering}] which  argues that the effect is too small and proposes another mechanism related to the modulation of electron-electron scattering in the presence of a magnetic field.

\section{Longitudinal $(2 k_F,0)$ modulation}
\label{sect:longitudinal}

\begin{figure}[h!]
\begin{center}
{\epsfxsize 4cm \epsffile{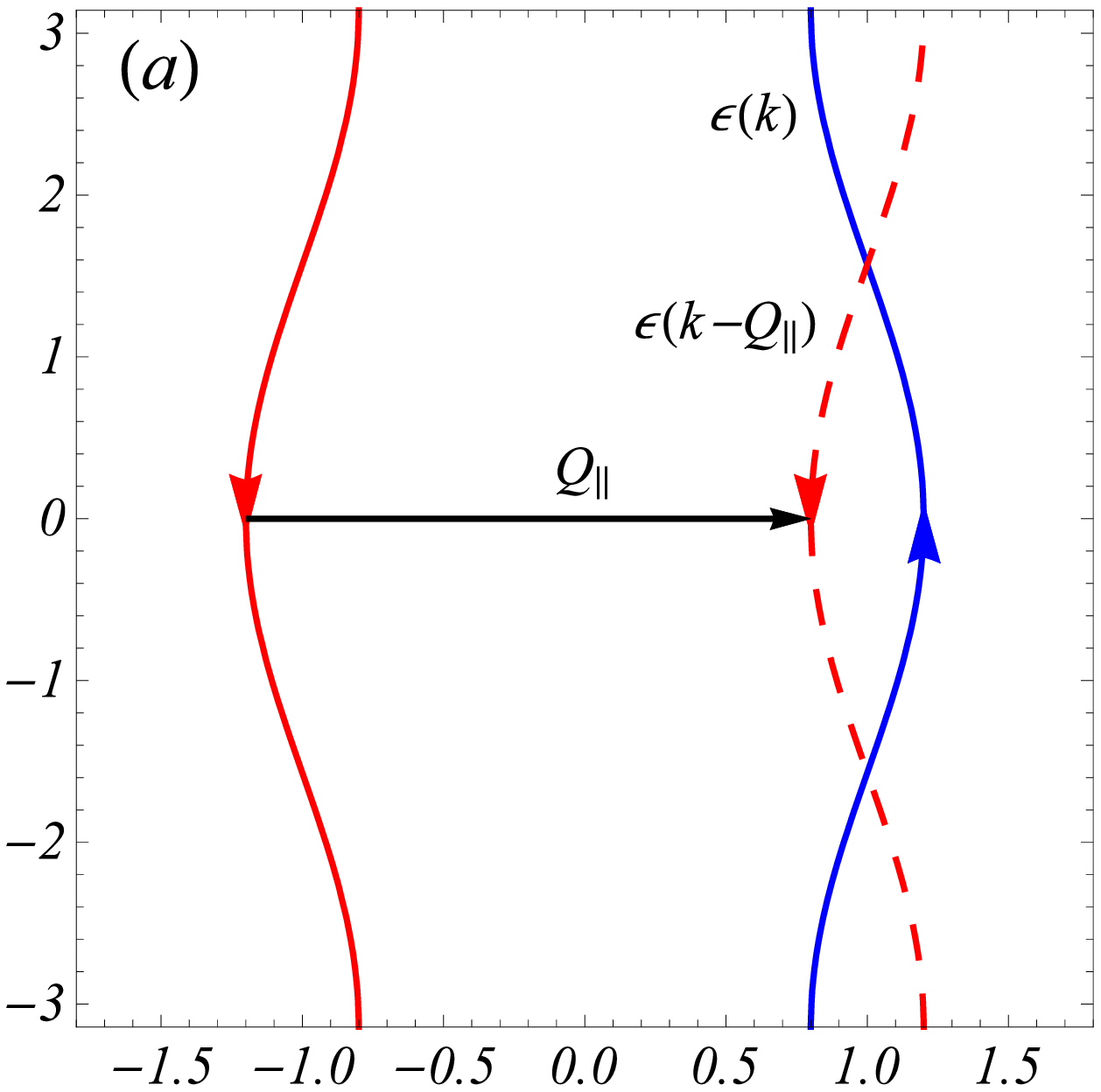}}
{\epsfxsize 4cm \epsffile{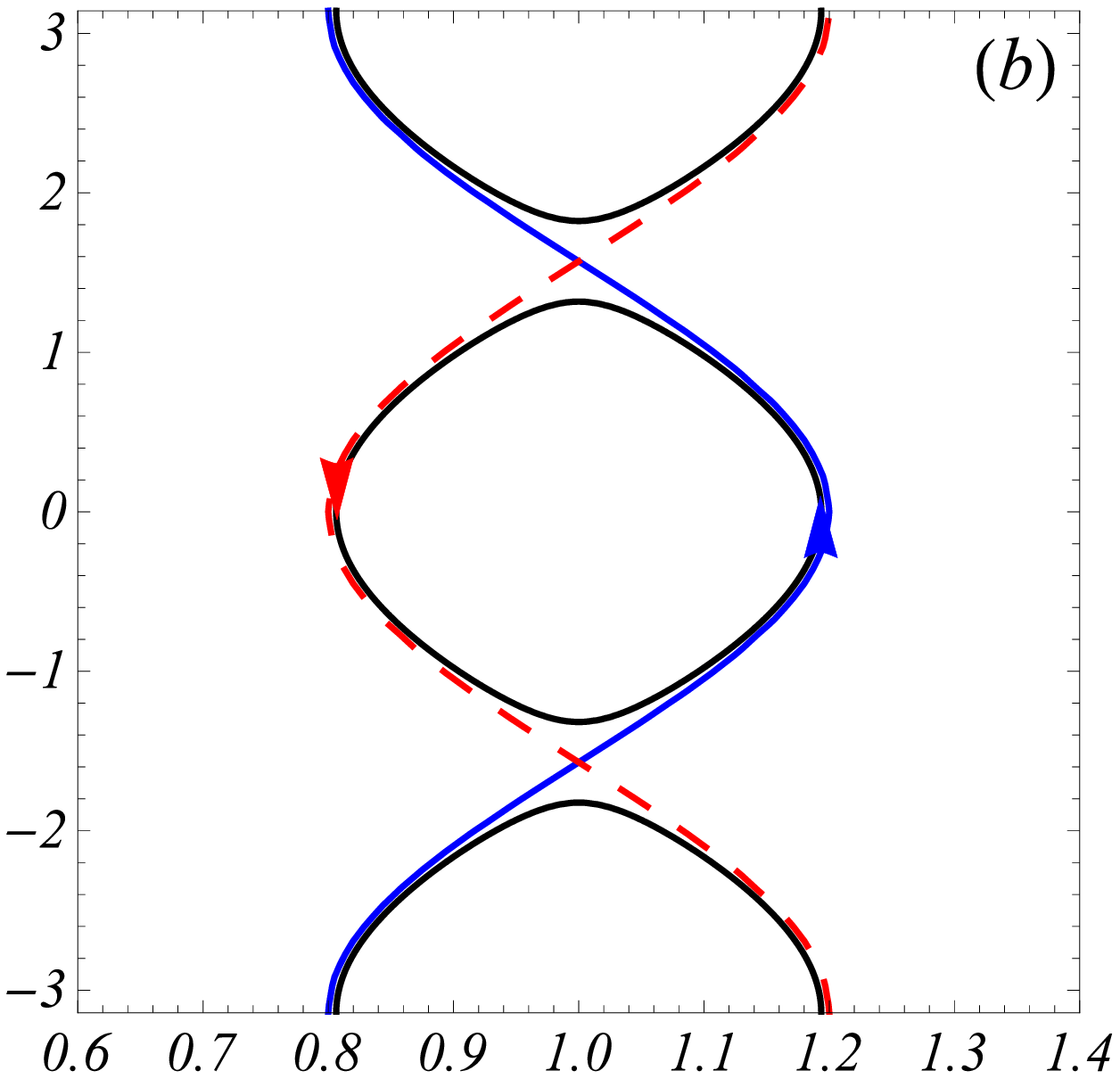}}
\caption{\small \it a) A modulation at a parallel wave vector $\Q_\parallel=(2 k_F,0)$ couples opposite sides of the FS. In a magnetic field, the coupled trajectories have opposite signs (blue and dashed red). b) The opening of a gap at this wavevector creates  electron and hole  closed pockets (black), whose  motion is quantized, leading to   SdH oscillations with frequency $t_b$.  The magnetic breakdown broadens Landau levels. }
\label{fig_Qparallel}
\end{center}
\end{figure}

We now consider the existence of a longitudinal modulation with amplitude $\Delta_\parallel$ at wave vector $\Q_\parallel=(2 k_F ,0)$, as it   exists in  (TMTSF)$_2$NO$_3$ under pressure.\cite{Kang95,Vignolles05}
This case is quite different from the previous one, since the modulation vector $\Q_\parallel=(2 k_F,0)$ couples states on {\it opposite} sheets of the Fermi surface (Fig. \ref{fig_Qparallel}-a).    The corresponding Hamiltonian has the form~:
\be{\cal H}_\parallel(\k)= \left(
  \begin{array}{cc}
    \ep_\k & \Delta_\parallel \\
    \Delta_\parallel & \ep_{\k-\Q_\parallel} \\
  \end{array}
\right) \ee
with
\be \ep_{\k - \Q_\parallel} = - \hbar v_F ( k_x - k_F) - 2 t_b \cos k_y b - 2 t_b' \cos 2 k_y b  \label{shiftedQpara}  \ee
to be compared with (\ref{shiftedQA}) where the modulation was at wave vector $\Q_\perp=(0,\pi/b)$. Like in the previous case, the $t'_b$ term does not play an important role here, and we  set $t'_b=0$. The spectrum is given by
\be E_\k= -2 t_b  \cos( k_y b)  \pm \sqrt{ \Delta^2 + \hbar^2 v_F^2(k_x-k_F)^2 } \ee
and the equation of the FS ($E_\k=0$) is~:
\be   k_x  =  k_F     \pm {1 \over \hbar v_F} \sqrt{ 4 t_b^2 \cos^2 k_y b -  \Delta_\parallel^2  }   \ee
 to be contrasted with (\ref{FSA}) for the $\Q_\perp$ modulation. The Fermi surface is shown on Fig.  \ref{fig_Qparallel}-b.   It    defines electron and holes pockets of equal size,  leading to {\it closed} orbits in a magnetic field.\cite{Fortin2009} The area of these orbits is ${\cal A}= {8 t_b \over \hbar v_F b}$ leading to Shubnikov-de Haas oscillations with a characteristic field)~:

\be \BBS(\Delta_\parallel) = {4 t_b \over \pi e v_F b}\  F_\parallel(\Delta_\parallel /2 t_b)   \ ,  \label{Bstarparallel} \ee
with
\be
F_\parallel(x)= \int_0^{\arccos x} \sqrt{\cos^2 t - x^2}\,  dt \ .  \ee
For a small gap, this is the same characteristic field as in the previous case. However, it {\it decreases}
 when the gap increases, since the size of the closed pockets decreases. The variation is shown on Fig. \ref{fig:freqs}.
  It is indeed interesting to contrast the gap ($\Delta_\parallel$) dependence of the characteristic field of these SdH oscillations, with the gap ($\Delta_\perp$) dependence of the characteristic field of the Stark oscillations.

 Due to magnetic breakdown, there is a finite probability of tunneling between the closed orbits shown on  Fig.  \ref{fig_Qparallel}-b, leading to  open orbits. To calculate this probability, we expand the Hamiltonian  near a crossing point   $k_y=\xi \pi/(2b) + q_y$ with $\xi=\pm 1$. It takes the  form

\be{\cal H}_\parallel=   \left(
  \begin{array}{cc}
   \delta   + 2  \xi t_b b  \, q_y  & \Delta_\parallel \\
    \Delta_\parallel & -\delta  +2  \xi t_b b \,  q_y  \\
  \end{array}
\right) \ee
with the spectrum ($\delta= v_F (k_x - k_F ) $)
\be E_\k=   2 \xi t_b b q_y \pm \sqrt{ \delta^2 +\Delta_\parallel^2  }  \ . \ee
In a magnetic field, there is a   key difference with the previous case since the motion is {\it opposite} along the two sheets of the Fermi surface :
\be \hbar q_y = \pm F t = e v_F \BB t \ee
with $\hbar v_y =2 t_b  b$. Therefore the time dependent Hamiltonian reads

\be{\cal H}_\parallel(t)=  \left(
  \begin{array}{cc}
   \xi v_y  F t   & \Delta_\parallel\\
    \Delta_\parallel &  -\xi  v_y  F t  \\
  \end{array}
\right) \ee
and the LZ probability  to tunnel from one closed pocket to another  is given by

\begin{equation}
p_\parallel= e^{\displaystyle -{\BB_\parallel \over \BB} }  \quad     \mbox{with} \quad  \BB_\parallel=
 \pi  {\Delta_\parallel^2   \over  e \hbar  v_y  v_F }= {\pi \over 2}{ \Delta_\parallel^2 \over e t_b  b v_F } \ .
\label{MB2} \end{equation}
 This magnetic breakdown leads to a broadening of the Landau levels $E_n =(n+1/2) e v_F b \BB$, which has
 been estimated   by different methods  in Refs.[\onlinecite{Gvozdikov,Fortin2009}].
This broadening leads to a modulation of the SdH oscillations.

\section{$(2 k_F, \pi / b)$ Nesting ordering}
\label{sect:nesting}

\begin{figure}[h!]
\begin{center}
{\epsfxsize 4cm \epsffile{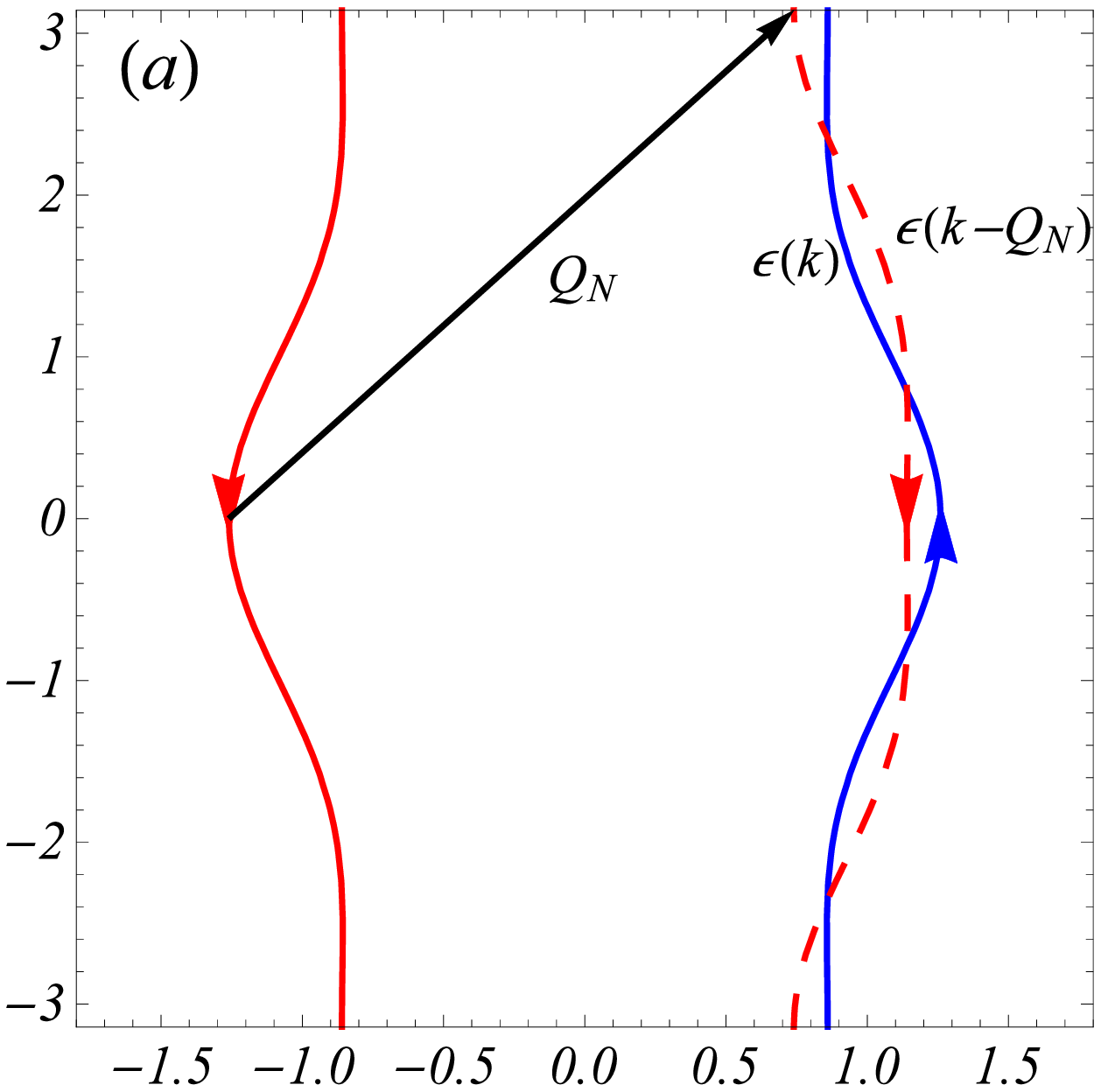}}
{   \epsfxsize 4cm \epsffile{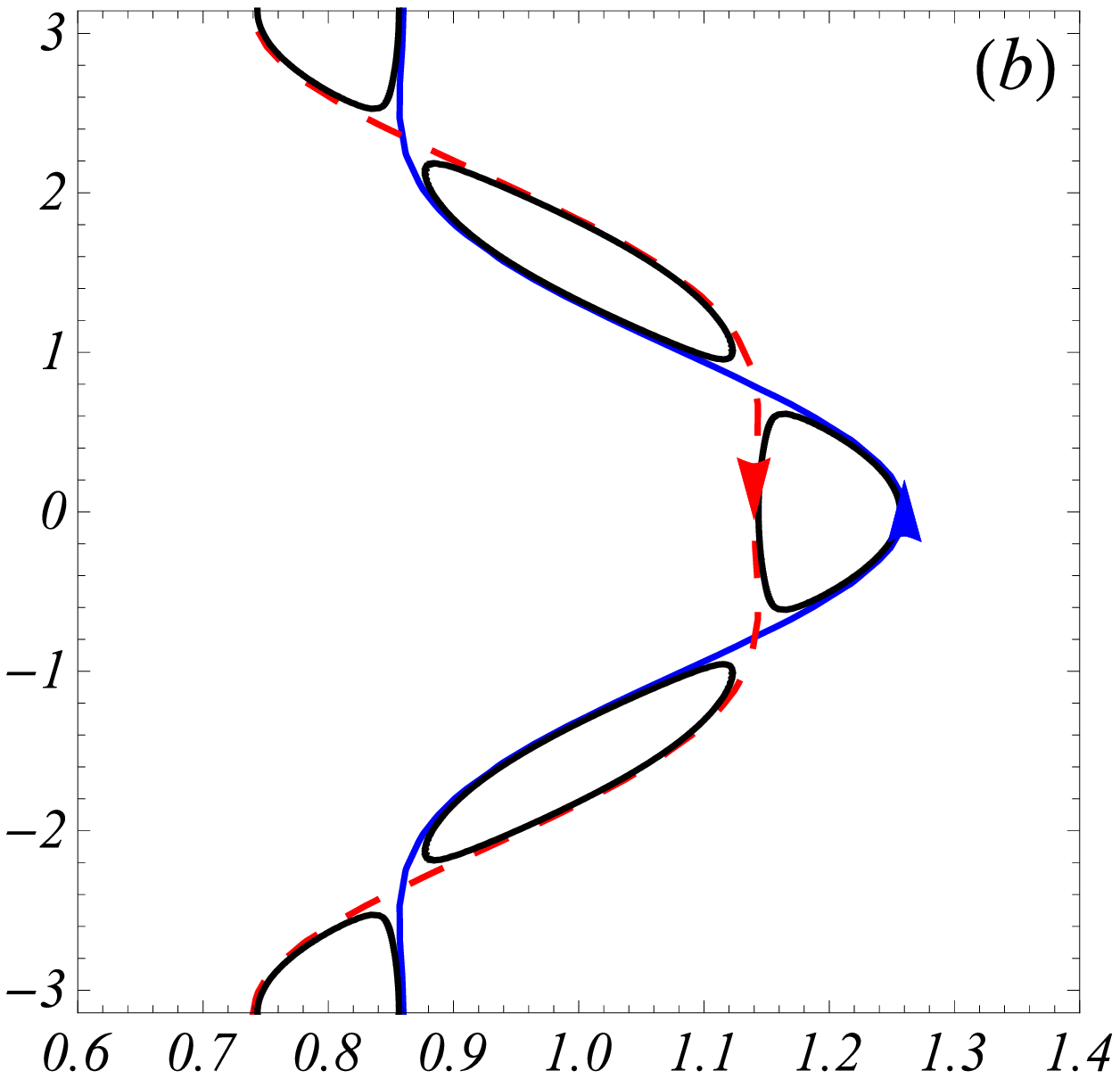}}
\caption{\small \it  a) A modulation at the nesting vector $\Q_N=(2 k_F,\pi/b)$ couples opposite sides of the FS. In a magnetic field, the coupled trajectories flow along opposite  directions (blue and red dashed). b) The opening of a gap at this wavevector creates electronic closed pockets (black), whose motion is quantized, leading to SdH oscillations with frequency proportional to $t'_b$.  }
\label{fig_QN}
\end{center}
\end{figure}

Like in the previous case, this modulation  couples states on {\it opposite} sheets of the Fermi surface (Fig. \ref{fig_QN}-a). The difference is that the nesting of the FS is almost perfect,\cite{remarknesting} and the characteristic field $\BBS$ of
 the SdH oscillations is set by the energy scale $t'_b$ instead of $t_b$.  The Hamiltonian reads~:
\be{\cal H}_N(\k)= \left(
  \begin{array}{cc}
    \ep_\k & \Delta_N \\
    \Delta_N & \ep_{\k-\Q_N} \\
  \end{array}
\right) \ee
with
\be \ep_{\k - \Q_N} = - \hbar v_F ( k_x - k_F) + 2 t_b \cos k_y b - 2 t_b' \cos 2 k_y b  \label{shiftedQN}  \ee
to be compared with (\ref{shiftedQA}). The spectrum is given by
\be E_\k= -2 t_b' \cos(2 k_y b) \pm \sqrt{ \Delta_N^2 + [\hbar v_F(k_x-k_F) -2 t_b \cos k_y b]^2 } \ee
and is shown on Fig. (\ref{fig_QN}-b).
The equation of the FS is

\be   k_x  =  k_F  - {2 t_b \over v_F} \cos k_y b    \pm {1 \over v_F} \sqrt{ 4 {t'_b}^2 \cos^2 2 k_y b -  \Delta_N^2  } \ .  \label{FSN} \ee

The Fermi surface consists in four inequivalent small electron pockets, leading to {\it closed} orbits in a magnetic field. However due to magnetic breakdown there is a finite probability of tunneling between these orbits, leading to larger or even open orbits. To calculate this probability, we expand the Hamiltonian  near a crossing point   $k_y=\xi \pi/(4b) + q_y$ with $\xi=\pm 1$. It takes the  form

\be{\cal H}_N(\k)=   \left(
  \begin{array}{cc}
   \delta - 4  \xi t_b' b  \, q_y  & \Delta_N \\
    \Delta_N & \delta -4 \xi t_b' b \,  q_y  \\
  \end{array}
\right) \ee
with the spectrum ($\delta= v_F (k_x - k_F ) - t_b \sqrt{2}$)
\be E_\k= - 4 \xi t_b' b q_y \pm \sqrt{ \delta^2 +\Delta_N^2  } \ .  \ee
In a magnetic field, the motion is {\it opposite} along the two sheets of the Fermi surface :
\be q_y = \pm F t = e v_F \BB t \ee
with $\hbar v_y'= 4 t_b' b$. Therefore the time dependent Hamiltonian reads

\be{\cal H}_N(t)=  \left(
  \begin{array}{cc}
  - \xi v_y' F t   & \Delta_N \\
    \Delta_N &  \xi  v_y' F t  \\
  \end{array}
\right)  \ .\ee
The LZ probability  to tunnel between closed orbits is  therefore given by

\begin{equation}
p_N= e^{\displaystyle -{\BB_N \over \BB} }  \quad     \mbox{with} \quad  \BB_N=\pi {\Delta_N^2 \over \hbar v_y' F }=  {\pi \over 4}{ \Delta_N^2 \over e t_b' b v_F } \ .
\label{MB3} \end{equation}

  This tunneling leads to a broadening of the Landau levels and to a modulation of the SdH oscillations.\cite{Gvozdikov}  These slow oscillations have  been not observed yet in  the SDW phase of  Q-(TMTSF)$_2$ClO$_4$ or   (TMTSF)$_2$PF$_6$.


\section{Nesting ordering and transverse modulation}
\label{sect:nesting-transverse}

\begin{figure}[h!]
\begin{center}
{\epsfxsize 4cm \epsffile{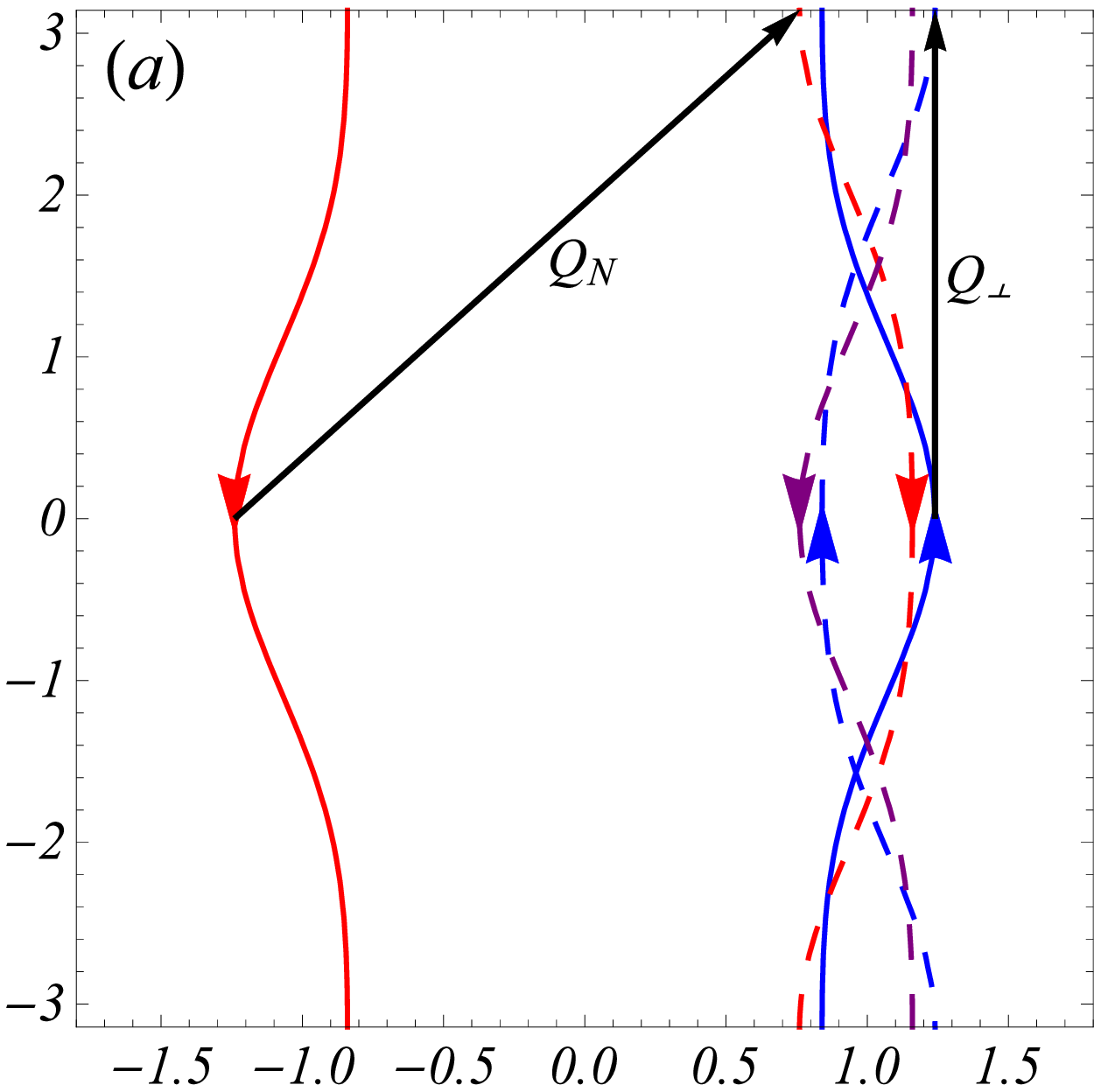}}
{   \epsfxsize 4cm \epsffile{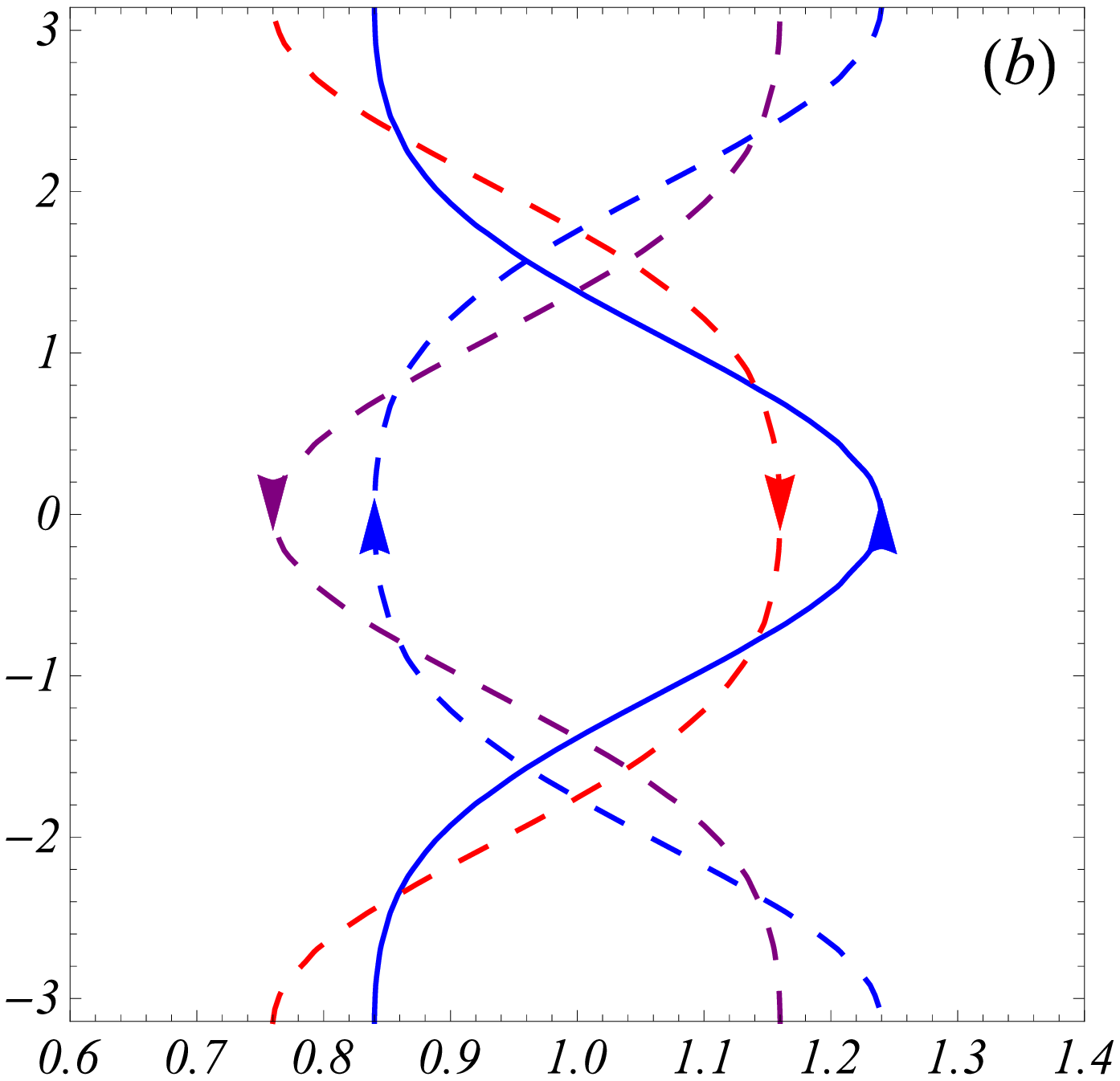}}
{   \epsfxsize 6cm \epsffile{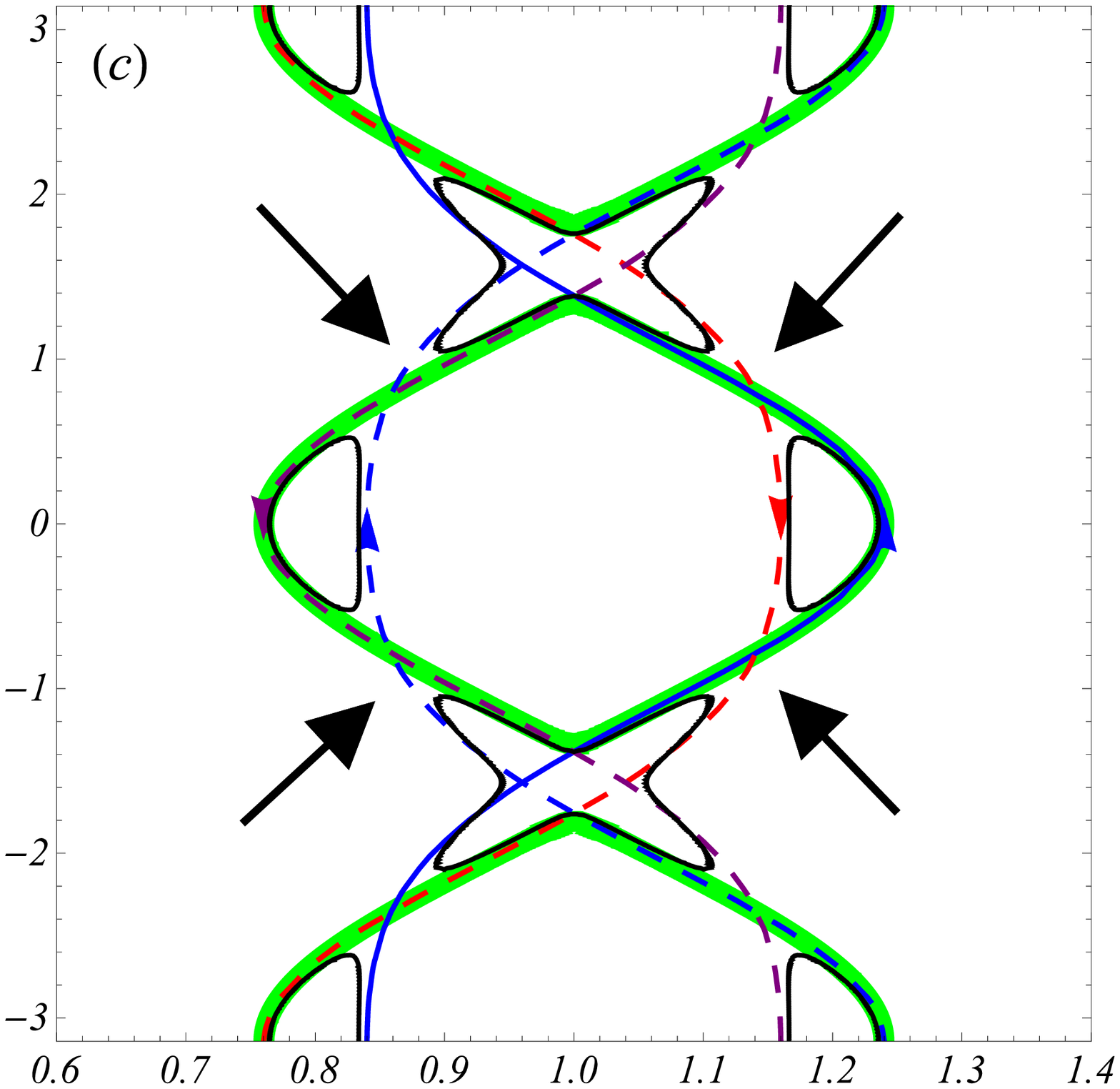}}
\caption{\small \it    a) The presence of  the two modulations $\Q_N=(2 k_F,\pi/b)$ and $\Q_\perp=(0,\pi/b)$ couples four states   ($\k$ blue, $\k-\Q_N$ dashed red, $\k-\Q_\perp$ dashed blue   and $\k-\Q_\perp + \Q_N$  dashed purple),
 creating six electronic closed pockets (black) (b) whose motion is quantized, leading to SdH oscillations with frequency $t'_b$. Notice the existence of new $X$-shape pockets.   Magnetic breakdown through the gaps (arrows) leads to large closed orbits (shown in green in the bottom figure c) whose size is proportional to $t_b$. }
\label{fig_QNperp}
\end{center}
\end{figure}

We consider now the case where two periodicities coexist, one at wave vector $\Q_N= (2 k_F, \pi/b)$ with amplitude $\Delta_N$, and one at wave vector $\Q_\perp= (0, \pi/b)$ with amplitude $\Delta_\perp$. This could be the case for example if a DW and a modulation due to anion ordering would coexist,   as in the anion ordered phase of (TMTSF)$_2$ClO$_4$. This situation has been studied extensively in Ref.[\onlinecite{Machida97}],
 and we present here a simple analytical description. It first has to be stressed that this situation {\it cannot} be reduced to the case of one modulation at wave vector $\Q_N - \Q_\perp$. This is a novel situation in which   {\it  four} states are coupled.
Indeed these two  modulations  couple a state with wavevector $\k$ to a infinity of states $\k+ m \Q_N  + n \Q_\perp$.
Considering states closed to the FS, one sees that a vector $\k$ on the right side of the FS is coupled to $\k - \Q_N$, $\k - \Q_\perp$ and
$\k - \Q_N+ \Q_\perp$. The energy spectrum $E_\k$ is therefore  given by the eigenvalues of the Hamiltonian~:

\be {\cal H}(\k)= \left(
  \begin{array}{cccc}
     \epsilon_{\k} & \Delta_{\Q_N} & \Delta_{\Q_\perp} & 0 \\
   \Delta_{\Q_N^*} &\epsilon_{\k -\Q_N} &0 &\Delta_{\Q_\perp} \\
   \Delta_{\Q_\perp^*} & 0
 &\epsilon_{\k -\Q_\perp} & \Delta_{\Q_N} \\
   0 &\Delta_{\Q_\perp^*} &\Delta_{\Q_N^*} & \epsilon_{\k -\Q_N+\Q_\perp} \\
  \end{array}
\right) \ . \label{HNperp}
\ee

Fig.\ref{fig_QNperp}-a shows the four states which are coupled by the modulations, with opening of a gap near $k_y= \pm \pi/(4b)$ and $k_y= \pm 3 \pi /(4b)$, as seen on Fig. \ref{fig_QNperp}-b. The new spectrum exhibits two kinds of electronic pockets~:  pockets with a banana shape as in the case of pure DW modulation and new pockets with a X-shape. Quantization of these orbits in a field should lead to oscillations with frequency $t'_b$. The important new feature here is the possibility of magnetic breakdown through four gaps, leading to closed orbits of large size, related to $t_b$ (Fig. \ref{fig_QNperp}-c) leading therefore to rapid quantum oscillations. The probability of having such large orbits involves four magnetic breakdowns (shown by arrows in Fig. \ref{fig_QNperp}-c with probability $p_N$ calculated above (Eq. \ref{MB3}) and two Bragg reflections. The
amplitude associated to the Bragg reflection may be more difficult to calculate since it involves four waves instead of two.   This scenario is therefore expected to lead to rapid oscillations superimposed to the slow oscillations. It naturally explains  the existence of thermodynamic rapid oscillations in the FISDW of R-(TMTSF)$_2$ClO$_4$,\cite{Kang91b,Uji96}   as well as in (TMTSF)$_2$ReO$_4$ under pressure\cite{Schwenk86,Kang91b}.

\section{Nesting ordering and longitudinal modulation}
\label{sect:nesting-longitudinal}

In this situation, two periodicities coexist, one corresponding to a DW ordering at wave vector $\Q_N=(2 k_F, \pi/b)$ and a modulation at the longitudinal wave vector $\Q_\parallel= (2 k_F, 0)$ as the one induced by the anion ordering in  (TMTSF)$_2$NO$_3$. As in the previous case, four states are coupled close to the Fermi level, $\k$, $\k- \Q_N$, $\k- \Q_\parallel$ and $\k- \Q_N+ \Q_\parallel$. This situation is therefore very similar to the previous one, since one has $\Q_\parallel= \Q_N - \Q_\perp$, and the Fermi pockets in this case resemble those presented in Fig.\ref{fig_QNperp}.\cite{Machida97}  One expects a superposition of rapid oscillations related to the large pockets due to magnetic breakdown between small pockets responsible for slow oscillations.
  This scenario naturally explains the two series of oscillations observed in the SDW phase of (TMTSF)$_2$NO$_3$.\cite{Audouard94,Kang95,Vignolles05}

\section{Two commensurate SDW}
\label{sect:nesting-commensurate}
\begin{figure}[h!]
\begin{center}
{\epsfxsize 4cm \epsffile{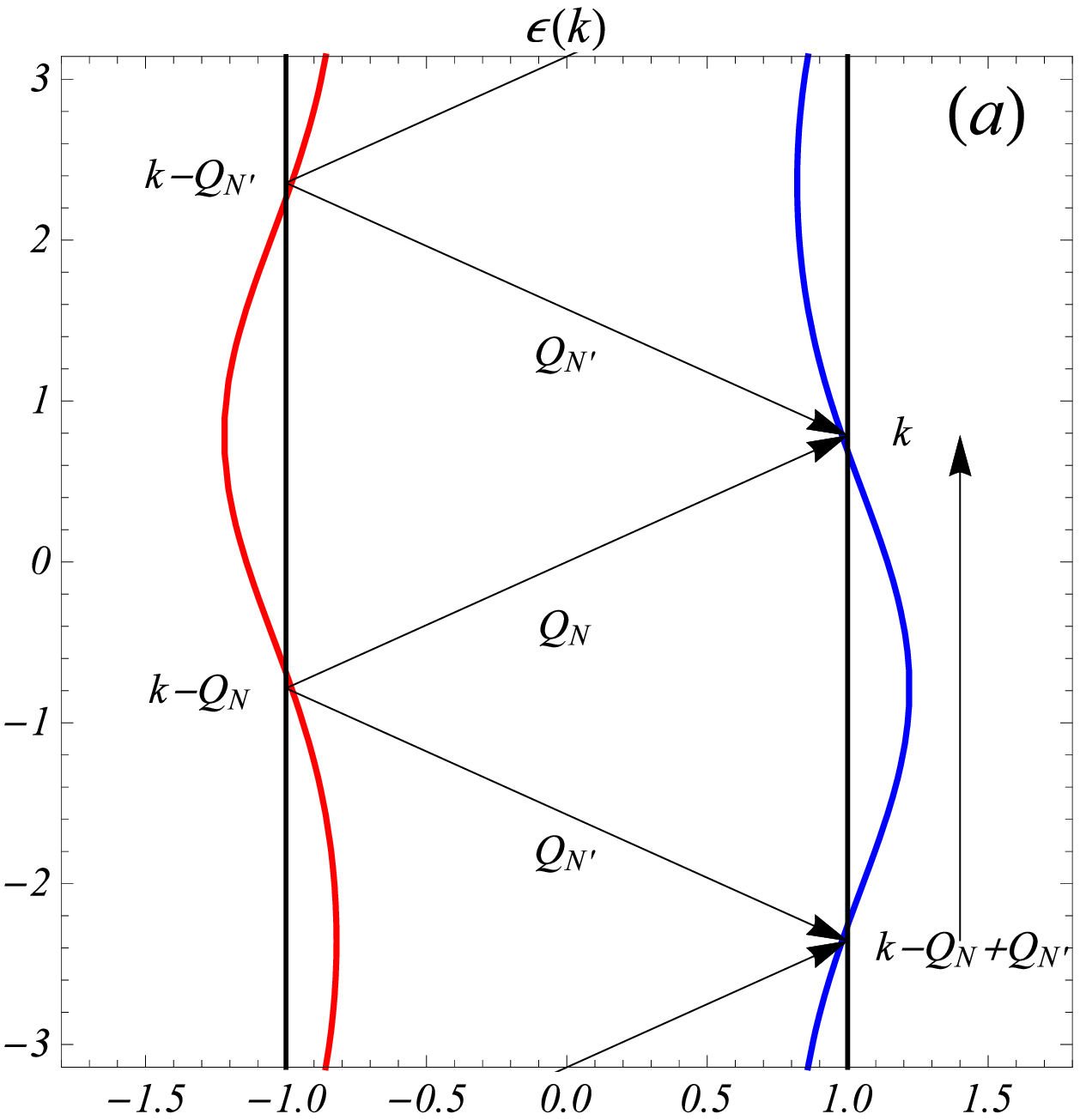}}
{   \epsfxsize 4cm \epsffile{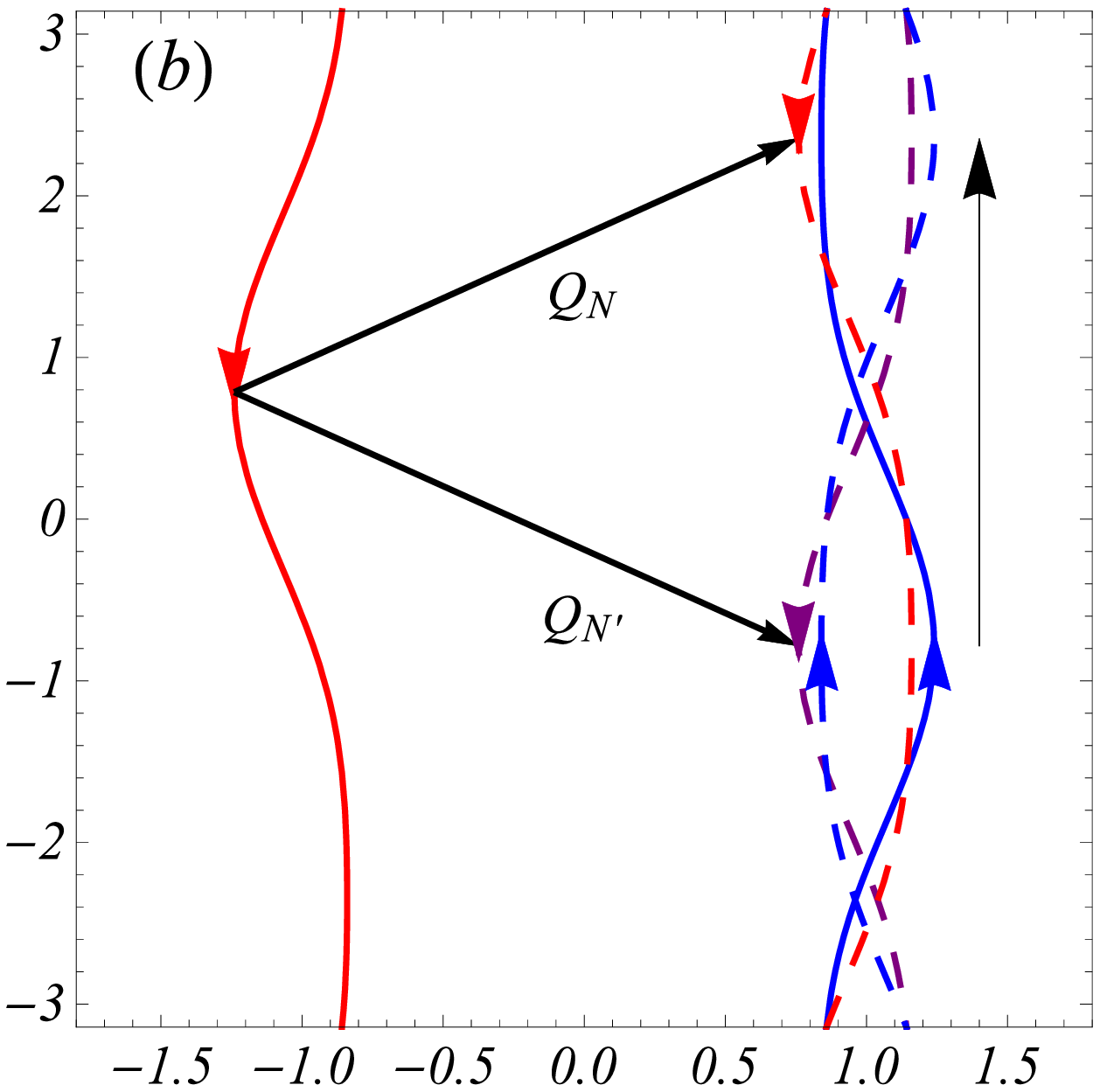}}
{   \epsfxsize 4cm \epsffile{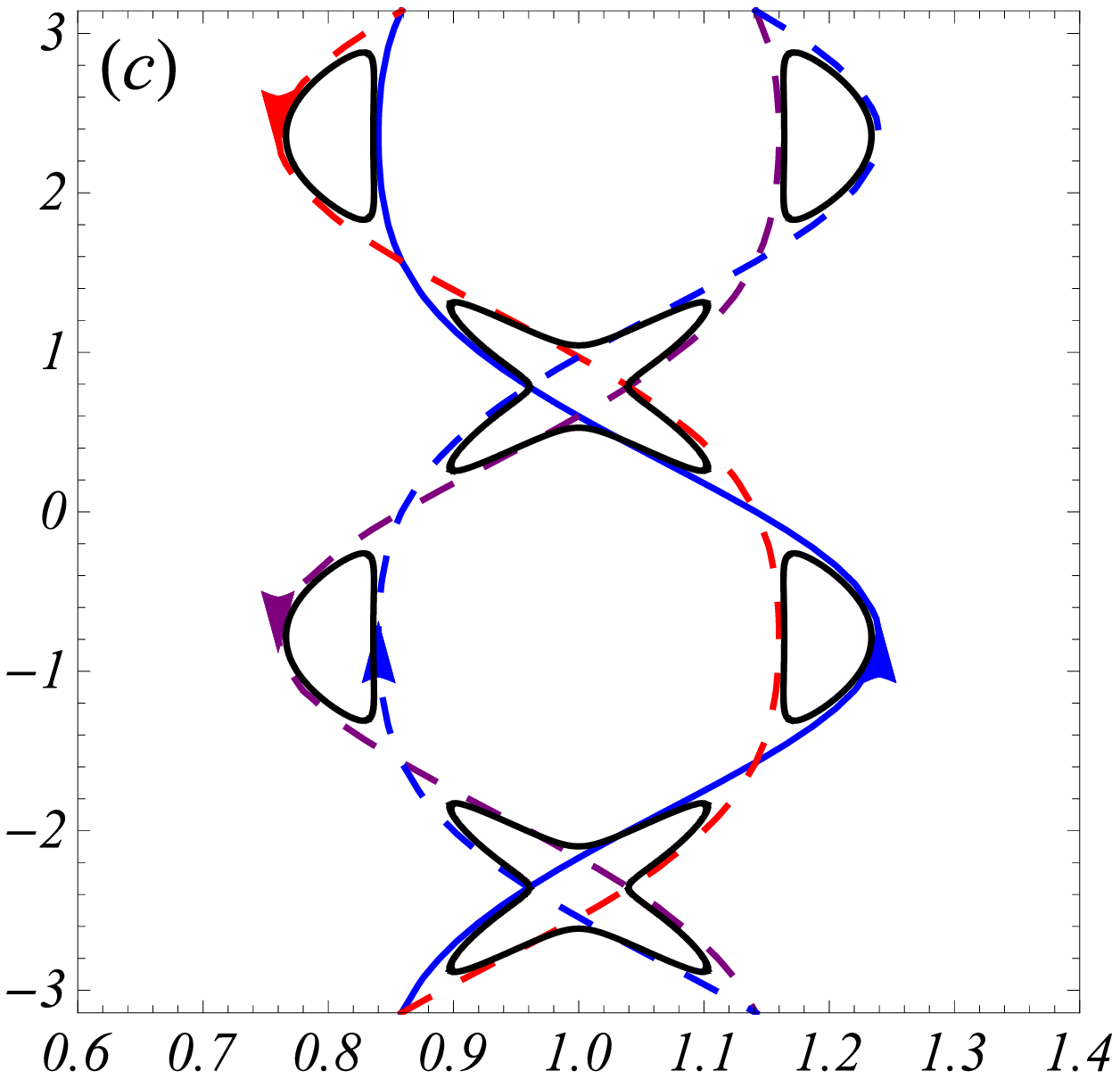}}
{   \epsfxsize 4cm \epsffile{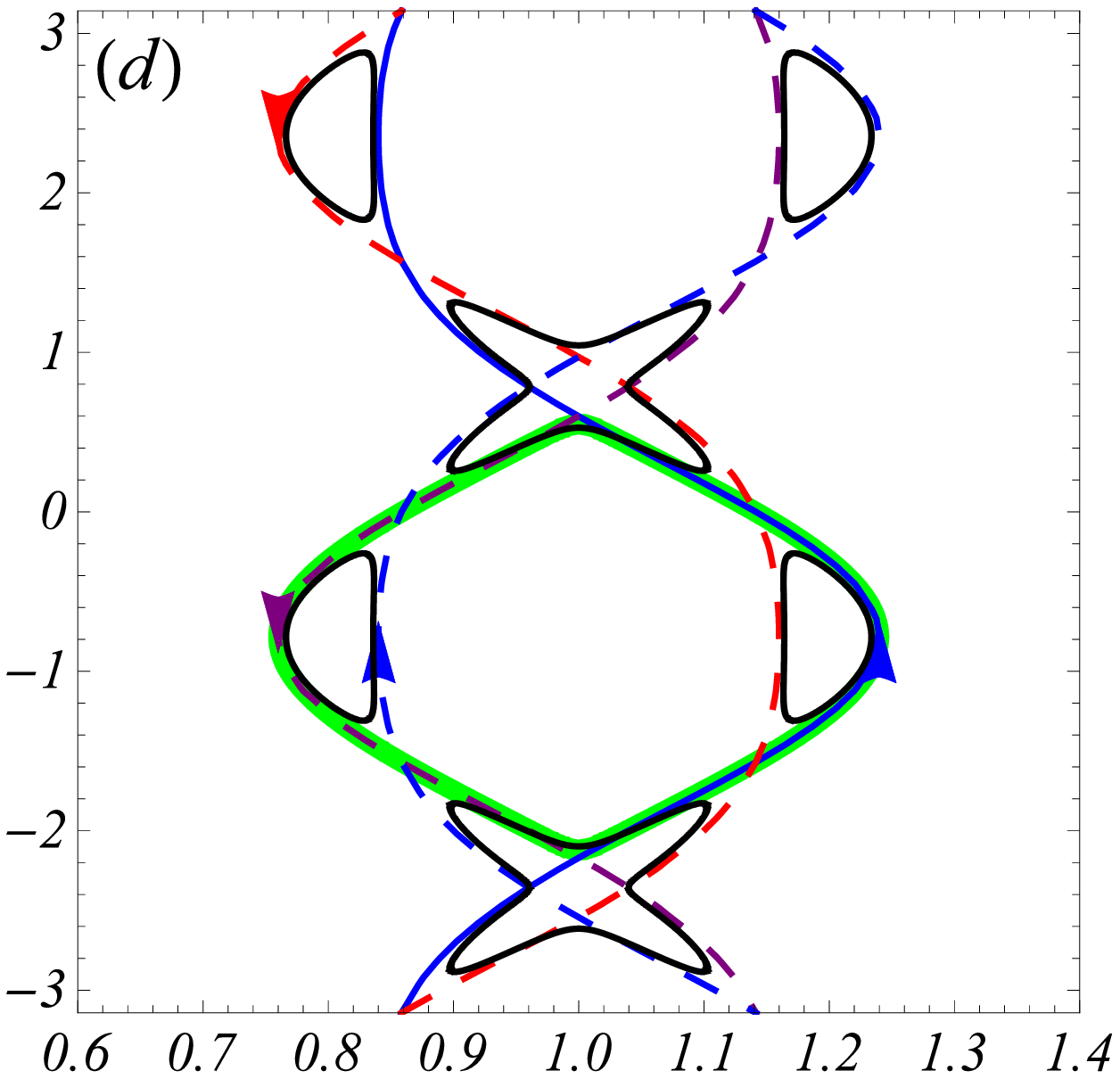}}
\caption{\small \it   Fermi surface related to the dispersion relation (\ref{DR2}), and modulation wave vectors $\Q_N=(2 k_F, \pi/2b)$ and $\Q_{N'}=(2 k_F, -\pi/2b)$. a) The presence of  these two modulations   couples four states,  couples four states   ($\k$ blue, $\k-\Q_N$ dashed red, $\k- \Q_{N'}$  dashed purple and $\k-\Q_N+\Q_{N'}$ dashed blue), creating six electronic closed pockets (b) whose motion is quantized, leading to SdH oscillations with frequency $t'_b$. Notice the existence of new $X$-shape pockets.   Magnetic breakdown through the gaps   leads to closed orbits (green curve in d) whose size is proportional to $t_b$.   }
\label{fig_Qtricl}
\end{center}
\end{figure}

  The two previous sections  show  interesting situations  when two modulations coexist. This is the  case in (TMTSF)$_2$ClO$_4$ and (TMTSF)$_2$NO$_3$,  where a DW modulation  coexists with a modulation due to anion ordering, respectively at wave vectors $\Q_\parallel=(0,\pi/b)$ and $\Q_\perp=(2 k_F, 0)$.
 However,  rapid oscillations may also  exist even in the absence of  anion ordering, as it is the case in (TMTSF)$_2$PF$_6$.
We now discuss another situation where two DW modulations can coexist, leading to a similar structure of the FS as in the two previous sections.
Instead of the orthorhombic symmetry considered until now, we consider the triclinic symmetry pertinent to Bechgaard salts. The FS is distorted such that the two sheets are not facing each other as in the previous cases, but one is shifted with respect to the other one (Fig. \ref{fig_Qtricl}-a). Here we keep the same simple orthorhombic model where the shift is described by a phase $\phi$~:\cite{Yamajibook}
\begin{eqnarray}
 \ep(\k)&=& \hbar v_F (|k_x| - k_F) - 2 t_b \cos[  k_y b + \phi ~\mbox{sgn}(k_x)] \\
 &-&2 t_b'\cos [ 2 k_y b  + 2 \phi ~\mbox{sgn}(k_x)]  \ . \label{DR2}  \end{eqnarray}
We consider the commensurate situation when the shift  $\phi= \pi/4$ as  shown on Fig.    \ref{fig_Qtricl}-a.
\medskip

  Assume the existence of a modulation close to perfect nesting at wave vector $\Q_N=(2 k_F, \pi/2 b)$, which couples states $\k$ and $\k - \Q_N$.
Consider also the modulation at wave vector  $\Q_{N'}=(2 k_F, -\pi/2 b)$ which does not provide a so good nesting. However, the state $\k - \Q_{N'}$ is coupled to the state $\k + \Q_{N'} - \Q_N$ which is close to the FS.
Therefore, we   have to consider the coupling between {\it  four} states near the Fermi level   $\k$, $\k-\Q_N$, $\k-\Q_{N'}$, $\k- \Q_N+ \Q_{N'}$.\cite{remarkcommensurate} The coupling vectors are   represented on Fig. \ref{fig_Qtricl}-b.

Let $\Delta_N$ and   $\Delta_{N'}  <  \Delta_N$ be the amplitude of these two modulations.
The new spectrum is obtained by diagonalization of the $4 \times 4$ matrix~:

\be {\cal H}(\k)= \left(
  \begin{array}{cccc}
     \epsilon_{\k} & \Delta_{N} & \Delta_{N'} & 0 \\
   \Delta_{N}^* &\epsilon_{\k - \Q_N} &0 &\Delta_{N'}^* \\
   \Delta_{N'}^* & 0
 &\epsilon_{\k -\Q_{N'}} & \Delta_N^* \\
   0 &\Delta_{N'} &\Delta_N & \epsilon_{\k -\Q_N+\Q_{N'}} \\
  \end{array}
\right) \ .
\label{HQNcomm}\ee

It appears clearly that this model is very similar to the   one discussed in   section \ref{sect:transverse},  $\Q_N-\Q_{N'}$ playing the same role as $\Q_\perp$  ~: four states are coupled and we have the correspondance (compare (\ref{HQNcomm}) with the Hamiltonian (\ref{HNperp}))~:

\begin{center}
 \begin{tabular}{ l c l }
                 $\Q_N=(2 k_F, \pi/b)$ & $\longleftrightarrow$   & $\Q_N=(2 k_F, \pi/2b)$ \\
                  $\Q_\perp=(0, \pi/b)$ & $\longleftrightarrow$  & $\Q_N-\Q_{N'}=(0, \pi/b)$\\
                $\Q_N-\Q_\perp=(2 k_F, 0)$ & $\longleftrightarrow$  & $\Q_{N'}=(0, -\pi/2b)$\\
                \end{tabular}
                \end{center}

Like in the two previous cases, there are two kinds of closed Fermi pockets, shown on Fig.\ref{fig_Qtricl}-c, whose size is typically related to $t'_b$  so that one should expect also slow oscillations. Magnetic breakdown through the gaps leads to large trajectories (shown in green in Fig.\ref{fig_Qtricl}-d), whose size is related to $t_b$ and which may lead to RO. This mechanism may  actually be strengthened by Umpklapp processes due to the commensurability condition $\Q_N + \Q_{N'}= (4 k_F,0)=(2 \pi/a, 0)$ as proposed by Lebed.\cite{Lebed91} This scenario has been proposed in a qualitative form in Ref. [\onlinecite{Uji2}]. It provides a satisfactory explanation for the RO observed in  (TMTSF)$_2$PF$_6$. In this salt, it has been found that the characteristic field $\BBS$ increases with pressure in a way which supports the proportionality between $\BBS$ and $t_b$.\cite{Kornilov07}
 The slow oscillations related to the small pockets may be difficult to observe. The RO being due to magnetic breakdown induced closed orbits are also expected in magnetization experiments, but have not been observed.\cite{Uji2}

\section{Summary and discussion}
 \label{sect:conclusion}

  Quantum oscillations in  quasi-1D conductors with an open Fermi surface  originate from a reconstruction of the FS due to  potentials induced by periodic modulations.  Their nature can be classified by their characteristic frequency -- or characteristic magnetic field $\BBS$. The frequency of the rapid oscillations is typically related to the  energy $t_b$ which describes the warping of the FS. The slow oscillations are related to the energy scale $t'_b$ which measures the deviation from the sinusoidal warping. This paper is a review of  different mechanisms which may lead to the RO. The table (\ref{table1}) summarizes such mechanisms which may explain different types of RO in the Bechgaard salts.

 \begin{figure}[h!]
  \begin{center}
 \begin{tabular}{|c|c|c|}
   \hline
     & metallic & SDW \\
     \hline
 R-ClO$_4$/ReO$_4$  & $\Q_\perp$ (Stark, III) & $\Q_\perp$\! and\! $\Q_{N}$\! (VI) \\
 NO$_3$ low pressure  &     $\Q_\parallel$(IV)    &   $\Q_\parallel$\! and\! $\Q_{N}$\!  (VII)    \\
 NO$_3$ high pressure   &       $\Q_\perp$(III)\!     &   $\Q_\perp$\! and\! $\Q_{N}$\!  (VI)    \\
 PF$_6$/AsF$_6$/Q-ClO$_4$&  & $\Q_N$\! and\! $\Q_{N'}$\! (VIII) \\
   \hline
 \end{tabular}
 \end{center}
 \caption{  Possible scenarios to explain the fast oscillations in the TMTSF salts. We refer to the coupling vectors associated to each scenario and the corresponding section in the paper. Here ReO$_4$ is under pressure.  For    NO$_3$, low  and high pressure refer respectively to pressures  lower and higher than $7$ kbar.}
 \label{table1}
 \end{figure}

$\bullet$ The simplest mechanisms, described in  sections \ref{sect:transverse} and \ref{sect:longitudinal}, involve only one periodic modulation modulation.

A transverse modulation ($\Delta_\perp$) at wave vector  $(2 k_F, \pi/b)$ induces a pair of trajectories on the same side of the FS which may interfere via MB, and realize St\"uckelberg oscillations which, in the context of magnetic oscillations, have first been proposed by Stark and Friedberg.\cite{Stark,Uji96,Uji97} This modulation is induced by anion ordering in
 R-\cl \ at ambient pressure\cite{Yan87,Kang91b,Uji96}   or  in  \reo4 under pressure.\cite{Schwenk86}
 The characteristic fields of the oscillations are respectively $\BBS= 265$ T and  $\BBS=290$ T.
Due to this particular interference mechanism, these oscillations are observed solely on the conductivity but not on  thermodynamic properties.

 A longitudinal modulation ($\Delta_\parallel$) at  wave vector $(2 k_F, 0)$ couples states  on opposite side and the reconstructed FS has closed Fermi pockets, the size of which is proportional to $t_b$.
   This is the case in  \no \ at ambient pressure,
  with the anion ordering yielding a potential at wave vector $(2 k_F,0)$.\cite{Kang95,Vignolles05}

The evolution of the characteristic field $\BBS$ with the gap induced by the modulation may be easily calculated in these two previous scenarios  (sections \ref{sect:transverse}, \ref{sect:longitudinal}). In the first case, it increases with the amplitude of the gap  $\Delta_\perp$  while it decreases with the gap  $\Delta_\parallel$ in the second case (Fig. \ref{fig:freqs}).
This is consistent  with the observation of a smaller frequency in \no \ at ambient pressure ($\BBS=248$ T)\cite{Kang95,Vignolles05} compared to the cases of \cl \ ($\BBS=255-265$ T)\cite{Brooks99,Yan87,Chaikin83} or \reo4 \ ($\BBS=320-330$ T) under pressure.\cite{Schwenk86,Kang95,Vignolles05}

$\bullet$ More complex scenarios involve the superposition of two modulations, that is two order parameters with probably different amplitudes, a nesting modulation correlated to the apparition of a DW plus one of the two above-mentioned modulations related to an anion ordering.
The reconstruction of the FS in these cases is quite rich. It exhibits closed FS with characteristic size related to size $t'_b$ plus large trajectories induced by MB, the size of which is related to $t_b$. Therefore one expects a superposition of slow and rapid oscillations. Their respective amplitude may be difficult to predict theoretically, since it depends  on the relative amplitude of the two order parameters and on the MB probability.
This is the situation in R-ClO$_4$ where indeed RO have been observed in the FISDW.  It is noticeable that the RO in the metallic phase are seen only in transport experiments while, in the FISDW phases, they are seen both in transport and thermodynamic measurement. This proves that the mechanism for RO is different in the presence of a single transverse modulation or in the presence of two modulations (transverse + SDW), as already noticed in Ref.\onlinecite{Kang91b}.

$\bullet $ \tmno3 is a particularly rich  system. Rapid and slow oscillations are observed simultaneously in the low temperature and low pressure SDW phase,\cite{Audouard94} compatible with  the superposition of a nesting modulation plus an anion-ordering induced modulation at wave vector $(2 k_F, 0)$. The observed increase of the frequencies of both oscillation sequences is
compatible with an increase of the characteristic  energies $t_b$ and $t'_b$ under pressure.\cite{Vignolles05}

 However, a surprising feature is the sharp drop of the characteristic  frequency of the RO at a pressure of order 7 kbar from $325$ T to    $215$ T, together with a vanishing of the slow oscillations. Such a jump is hardly understandable without an important structural change. Several arguments suggest an anion order evolving from $\Q_\parallel=(2 k_F, 0)$ to $\Q_\perp=(0, \pi/b)$ around $7$ kbar.
 Firstly,  the shape of the transport anomaly passing through the anion ordering changes drastically.\cite{Mazaud80,Kang90}
  Secondly,  a recent  investigation of the angular dependent magnetoresistance of \tmno3 under 8.7 kbar has concluded to the existence of a Q1D Fermi surface even in the presence of anion ordering.\cite{Kang09}
   Thirdly, an ab-initio calculation is supporting these experimental findings.\cite{Alemany14}
  These arguments suggest a pressure induced phase transition from an anion induced modulation  at wave vector  $(2 k_F,0)$ and a modulation at wave vector $(0, \pi/b)$.
Therefore, the RO reported at $\approx$ 8 kbar\cite{Kang95,Vignolles05} might be actually related to Stark interferences between  two sheets of folded Fermi surfaces like in R-\tmc. The possibility of a pressure induced structural transition could deserve further experimental and theoretical investigations.

$\bullet$ The existence of   RO occurring in a SDW phase  without any role played by anions, like in \pf, \as  \ or   Q-\cl, is more problematic since a pure SDW ordering is expected to induce only slow oscillations with frequency related to $t'_b$.   The mechanism at the origin of such RO may find its origin in the  particular (triclinic) lattice symmetry of the Bechgaard salts.  As a consequence of this triclinicity,   the two sheets of the FS are actually shifted with respect to each other by  a vector close to the commensurate value $(0,\pi/2b)$ so that naturally in the presence of a SDW two modulations at wave vectors $(2 k_F , \pm \pi/2b)$ may coexist.\cite{Uji2,Lebed91} This situation is not so different from those described earlier where a SDW and anion induced modulation coexist. We argue that this may explain the existence of RO in  (TMTSF)$_2$PF$_6$ or Q-(TMTSF)$_2$ClO$_4$.
This scenario assumes the existence of two commensurate SDWs,   at wave vectors $(2 k_F, \pm \pi/2b)$,  so that the coupling between relevant states at the Fermi level is described by a $4 \times 4$ matrix. One may wonder if this commensurate model is a good approximation for the actual incommensurate situation. This is probably the case and given that the actual phase is close to commensurate, one expects that the SDW will exhibit discommensurations.

\medskip
  We close this conclusion by a discussion on open questions related to the commensurability of the SDW ordering in  several Bechgaard salts.

Several experimental results provide hints for a commensurate SDW state in (TMTSF)$_2$PF$_6$.  Strain and defect-free samples exhibit a  first order transition  with a sharp increase in the resistivity at $T_{SDW} \simeq 12$K. In addition, a non-linear conduction  is the signature of a pinning mechanism due to fourfold commensurability \cite{Kang90}. A weakly first order transition is also the conclusion of a study of the $^{13}$C NMR linewidth through the transition.\cite{Clark01} Moreover signs of commensurability are also clearly provided by $^{13}$C NMR spectra.\cite{Nagata13}
The examination of the $^{1}$H-NMR lineshape has led to a determination   SDW wave vector as $(\pi/a, \simeq \pi/2b)$  with a rather broad error bar for the transverse component.\cite{Delrieu86,Takahashi86b} This is consistent with a  {\it triclinic} symmetry, since the best nesting vector  for a simplified {\it orthorhombic} model would be ($\pi/a,\pi/b$). However the typical double horn shape of the $^{13}$C-NMR spectra  makes the existence of an incommensurate modulation undisputable at least in the temperature interval between 12K and 4K.\cite{Barthel93b,Nagata13}

Another puzzle concerns
the  anomalous behaviour in the temperature dependence of several properties in the   ambient pressure  SDW phases of (TMTSF)$_2$X  salts.
The purest samples exhibit a significant enhancement of the resistivity at 4K but more important, the amplitude of RO  departs from the Lifshitz-Kosevich behaviour:~:  it has  a sudden drop at $T ^\star \approx 4$K without noticeable change of the frequency in \pf\ , \as \cite{Ulmet85} and  Q-\cl.\cite{Brooks99}
Interestingly, an anomalous behavior has also been noticed in the same temperature domain for the NMR properties namely, a sharp drop of proton\cite{Takahashi86a} and $^{77}$Se spin-lattice relaxation rates in \tmp6\cite{Valfells97} and Q-\cl\cite{Nomura93}. The quasi temperature independent relaxation rate below $T_{SDW}$ has been taken as an evidence for phason fluctuations governing the relaxation down to $T^\star$.\cite{Clark93,Wong93,Barthel93b}  Below that temperature the nuclear spin relaxation slows down and exhibits an activated behaviour suggesting the opening of a gap in the SDW phason modes.\cite{Takahashi86a,Valfells97} This feature is probably the signature of a fourfold commensurate SDW. An other clue in favour of commensurability is provided by microwave and radiofrequency measurements suggesting the existence below 4K of discommensurations close to commensurability.\cite{Zornoza05}

  This "4K anomaly" has not yet received any satisfactory interpretation. The exponential attenuation of the RO  below $T^\star$, without any modification of the characteristic frequency, has been interpreted as resulting from a suppression of the magnetic breakdown probability preventing the formation of large orbits.\cite{Brooks99} Such an hypothesis would require further ab-initio band structure calculations.

However, given the remarkable sharpness of the anomaly  observed by NMR\cite{Takahashi86a,Valfells97,Nagata13} towards a low temperature state characterized by an activation energy $\Delta=10$K,\cite{Takahashi87,Wong93} another hypothesis is that $T ^\star$ marks the existence of a real phase transition between an homogenous incommensurate SDW at high temperature and a low temperature state comprising inclusions of commensurate SDW domains within an incommensurate background.  Hence, the electronic scattering rate might be enhanced by the existence of domain walls occurring below $T^\star$.\cite{Nagata13} The RO  would retain the same oscillation frequency below $T^\star$ but the scattering rate could be strongly enhanced.\cite{Brooks99}

These problems have been unsolved for about thirty years and still raise interesting open questions.

\acknowledgments
  We are endebted to Prof. Jacques Friedel for his constant support of the research on organic conductors and his numerous discussions  and advices. D. J. acknowledges  stimulating discussions with Prof. L. P. Gor'kov on the subject of quantum oscillations. We also thank A. Audouard for useful comments and E. Canadell for fruitful discussions.


\begin{thebibliography}{99}
\bibitem{remarknesting}  The best nesting vector is actually slighlty different from this commensurate value.
\bibitem{remarkcommensurate}  This is correct because $\Q_N$ and $\Q_{N'}$ obey the commensurability relation $\Q_N - \Q_{N'}= (0, \pi/b)$.
\end{thebibliography}
 \end{document}